\documentclass[onecolumn]{aastex63}
\usepackage[utf8]{inputenc}

\submitjournal{BAAS}

\shorttitle{Socio-demographic study of the exoplanet direct imaging community 2: Follow-up}
\shortauthors{Leboulleux, Desai, Echeverri, et al.}

\graphicspath{{./}{figures/}}

\begin{document}

\title{A socio-demographic study of the exoplanet direct imaging community 2 
}

\correspondingauthor{Lucie Leboulleux}
\email{lucie.leboulleux@univ-grenoble-alpes.fr}

\author{Lucie Leboulleux}
\affiliation{Univ. Grenoble Alpes, CNRS, IPAG, 38000 Grenoble, France}

\author{Niyati Desai}
\affiliation{Department of Astronomy, California Institute of Technology, 1200 East California Blvd., Pasadena, CA, 91125, USA}

\author{Daniel Echeverri}
\affiliation{Department of Astronomy, California Institute of Technology, 1200 East California Blvd., Pasadena, CA, 91125, USA}

\author{Evangelia Kleisioti}
\affiliation{Leiden Observatory, Leiden University, P.O. Box 9513, 2300 RA Leiden, The Netherlands}
\affiliation{Faculty of Aerospace Engineering, TU Delft, Building 62 Kluyverweg 1, 2629 HS Delft, The Netherlands}

\author{Lorenzo K\"onig}
\affiliation{Department of Astronomy, California Institute of Technology, 1200 East California Blvd., Pasadena, CA, 91125, USA}

\author{Mathilde Malin}
\affiliation{LIRA, Observatoire de Paris, Université PSL, Sorbonne Université, Université Paris Cité, CY Cergy Paris Université, CNRS, 92190 Meudon, France}

\author{Elisabeth Matthews}
\affiliation{Max-Planck-Institut für Astronomie, Königstuhl 17, D-69117 Heidelberg, Germany}

\author{Saavidra Perera}
\affiliation{Department of Astronomy and Astrophysics, University of California San Diego, La Jolla, CA, 92093, USA}
\affiliation{Department of Astronomy and Astrophysics, University of California Santa Cruz, 1156 High Street, Santa Cruz, CA 95064}

\author{Schuyler Wolff}
\affiliation{Steward Observatory, University of Arizona, 933 North Cherry Avenue, Tucson, Arizona}

\author{Élodie Choquet}
\affiliation{Aix Marseille Univ, CNRS, CNES, LAM, Marseille, France}

\author{Elsa Huby}
\affiliation{LIRA, Observatoire de Paris, Université PSL, Sorbonne Université, Université Paris Cité, CY Cergy Paris Université, CNRS, 92190 Meudon, France}

\author{Garima Singh}
\affiliation{Gemini Observatory Northern Operations Center, 670 N. A'ohoku Place Hilo, HI 96720, USA}

\begin{abstract}

Recognizing and addressing under-representation, exclusion, and problematic behavior, including discrimination, within astronomy and astrophysics are crucial. Therefore, in 2019, a survey was conducted at the Spirit of Lyot conference to evaluate the socio-demographics and well-being of the exoplanet and disk imaging community. In this paper, we present the results of a second survey, conducted as a followup at the 2022 Spirit of Lyot conference held in Leiden, the Netherlands. Its objective was to 1) improve the evaluation of our community, both in terms of questions of interest and diversity, and 2) monitor the evolution of the metrics since the 2019 socio-demographic study. This new survey was sent to all participants in Spirit of Lyot and received $96$ responses, equivalent to a participation rate of $44\%$. It measured the visibility of respondents at conferences, their recognition through publications and projects, their experience of disrespect as victims or witnesses, their experience of inappropriate behaviors also as victims or witnesses, and their identification as allies of minorities in the field. These topics were studied with respect to job position, expatriation/non-expatriation, gender, belonging to another under-represented group (in terms of ethnicity, disability, or sexual orientation), and parenthood. Overall, comparing the survey respondent panel with the conference attendee information shows that survey respondents were more likely to be part of traditionally more-discriminated categories (women and non-permanent researchers, etc.). The survey also reveals biases against non-permanent researchers in terms of visibility (less invitation to Scientific Organizing Committees, less talks for PhD students). Regarding disrespect and inappropriate behaviors, the results are concerning: for instance, women and non-binary people are the most exposed to inappropriate behaviors, and are more likely to hold non-permanent positions. Considering these categories together, $36\%$ of female and non-binary PhD students (respectively $17\%$ of male PhD students) have experienced situations of inappropriate behaviors over the last $2.5$ years. There is also a clear trend between having experienced disrespect or inappropriate behavior and noticing that such a situation is happening. On a more positive note, most people ($82\%$), from all categories, are willing to learn about alliance. This paper also provides recommended practices to improve the well-being and visibility of under-represented groups in astronomy.

\end{abstract}


\section{Introduction} \label{sec:intro}

The topic of Inclusion, Diversity, Equity, and Accessibility (IDEA) in Science, Technology, Engineering, and Mathematics (STEM) academia has gathered attention through numerous studies over the last few years. The vast majority of articles point out the discrimination and violence faced by women in the field, ranging from differences in opportunities and sexist jokes to harassment. Among other issues, women in research are less credited than men with authorship \citep{Ross2022}, they are less promoted than men (e.g., with the French promotion system \citep{Berne2020}), they get less visibility through oral presentations at conferences (e.g., with the American Geophysical Union Fall Meeting \citep{Ford2018}), they are less awarded than men (e.g., with Nobel laureates \citep{Lunnemann2019}), or they receive negative biases in recommendation letters (ex. with geosciences \citep{Dutt2016}). Last but not least, the L'Oréal Foundation and IPSOS conducted a large survey in academia in 2023 including 117 countries and revealed that half of women have been victims of harassment at their workplace and two-thirds of them expressed a negative impact on their career (\href{https://www.fondationloreal.com/media/7706/download}{https://www.fondationloreal.com/media/7706/download}). As a consequence, women are more likely to leave academia than men: this phenomenon is called the ``leaky pipeline" \citep{Shaw2012}.

These barriers are not only faced by women. Parents and especially mothers also suffer biases and the so-called 'maternal wall' in academia (e.g., detrimental teaching evaluations \citep{Olabisi2021}, unequal access to permanent positions within French astronomy \citep{Berne2020}, and biases against parents among Brazilian scientists \citep{Staniscuaski2023}). The intersectionality of gender with other minority traits can further exacerbate these issues. For example, people of color, and particularly women of color, experience additional disadvantages (e.g., with teaching evaluations for Afro-descendant women in \citealt{Olabisi2021} and with career advancements in England and Wales for ethnic minority women in \citealt{Xiao2023}). Junior researchers, and in particular non-permanent researchers, are more vulnerable in the field due to career uncertainties and dependencies on their supervisors or collaborators. As an example, almost two-thirds of victims have experienced sexual harassment at the beginning of their career (\href{https://www.fondationloreal.com/media/7706/download}{see the IPSOS report concerning science researchers from 117 countries, page 61}). There is also a strong lack of accessibility for people with disabilities, with only $1\%$ of PhD candidates with disability among Science, Technology, Engineering, and Mathematics (STEM) doctorate students \citep{Aarnio2019, Moon2012}. Lesbian, Gay, Bisexual, Transgender, Queer (LGBTQ+) people are not exempted and also suffer from discriminations and toxic student or work environments (e.g., among UK scientists \citep{Tesh2019}, or in general physics and astronomy \citep{Ackerman2018}). Eventually, \cite{Berne2020} highlights the impact of the type of educational background ("elite" schools versus non-selective universities, which often correlate with students' socio-economic backgrounds) on access to permanent positions in French astronomy: the study finds that candidates from the "elite" education system are nearly three times more likely to obtain a permanent position in France compared to those who followed a university pathway. In addition to being part of a single group facing certain barriers, some individuals belong to multiple groups, and as a result, they experience the compounding effects of intersectionality, which can make their struggles even more complex \citep{Crenshaw1991, Carlone2007, Ong2018}.

The continuum of inappropriate behaviors, from biases and disparity to harassment and discrimination, negatively impacts the well-being of traditionally under-represented groups in STEM academia. Here we define 'well-being' as the ability to fully participate in the community, regardless of one's identities. Poor well-being, in turn, will drive members of under-represented groups away, therefore making the field even less diverse. Increased diversity, inclusion, and integration, however, generally allow for more innovation and improved performance \citep{Love2022, Smith-Doerr2017}. To counterbalance such negative effects, in addition to striving for a fairer society, it is crucial to measure these phenomena at various scales and within various fields of research, and put mitigation actions into place. 

The field of astronomy has already been the focus of several such studies \citep{Beri2023, Kewley2023, Hunt2021, Inno2021, Patat2020, Berne2020, Horstman2020, Richey2019, Primas2019, DOrgeville2014}. Conferences, and in particular repeating conferences, gather sub-groups of the astronomy community, and provide the opportunity to evaluate the status or well-being of this community regularly. Among other conferences, the Spirit of Lyot conference focuses on exoplanet and disks imaging and has so far been organized five times (2007, 2010, 2015, 2019, and 2022). The fourth conference (2019 in Tokyo, Japan) was the first to include a socio-demographic survey, leading to a publication focused on this community specifically \citep{Leboulleux2020}. The objective of the survey was to evaluate the demographics and social behavior with a focus on gender- and position-related differences. It mainly showed a lower visibility at conferences for early-career researchers (PhD students being under-represented at international conferences and postdocs being excluded from conference Science Organizing Committees), and inappropriate behaviors experienced by $33\%$ of respondents, and in particular by women. 

The fifth, and latest, edition of the Spirit of Lyot conference was held in Leiden, the Netherlands, in June 2022 with around $200$ participants. Before the conference, the organizing committee took the recommendation to do a follow-up of the first study and an improved version of the survey was released. The survey questions are shown in the Appendix \ref{sec:AppendixA}. This follow-up survey serves to monitor changes in the community since the first one. Additionally, even though this new survey was built upon the 2019 one, it was improved by changing some questions to reduce biases and rephrasing ambiguous or partial questions that are not ideal for the monitoring of our community. For instance, limited time frames were specified for when specific experiences occurred, and definition were provided for specific behaviors. In addition, more differentiation factors were included in addition to the previous ones of job position and gender. These additional factors include parenthood, under-representation (in terms of ethnicity, disability, and sexual orientation), and expatriation (meaning if one works and lives in a country that does not align with their nationality). 

It should also be noticed that between both surveys (2019 and 2022), the COVID-19 pandemic impacted the academic world largely, which was specifically measured by several studies \citep{Bohm2023, Leboulleux2021}. The pandemic has likely also affected the evolution of the metrics between both our surveys, such as publication rates or access to conferences or seminars.

Six main topics were addressed in the latest survey: 1) the demographics within respondents, 2) the visibility and exposure at conferences, 3) recognition with publications and projects, 4) situations of disrespect\footnote{In the survey, disrespect is described as "light behaviors that can make you or somebody uncomfortable, for instance cutting off somebody while they are talking, talking over another person without their consent, downplaying someone’s idea, expecting social role from younger people or women (taking notes, planning social events...)".}, both for victims and witnesses, 5) situations of inappropriate behaviors\footnote{In the survey, inapproriate behaviors are described as "any social behavior that can make you uncomfortable even if it is not necessarily legally reprehensible. Here are a few specific examples of inappropriate behaviors: condescending remarks, discriminating behavior based on ethnicity, inappropriate jokes, racist jokes, staring at, sexual remarks or questions at a work environment, disrespect based on one’s culture and identity, ignoring or excluding somebody during a meeting, preventing somebody from attending meetings, remarks on parental leaves, sexual harassment, bullying...".}, both for victims and witnesses, and 6) allyship. This last topic was fully absent from the previous study but was considered to be important during the set up of the survey since most people do not belong to under-represented groups (by definition). It should also be noted that despite disrespect, discrimination, inappropriate behaviors, and harassment being spread over a broader and continuous spectrum, we have only sampled them into two categories (disrespect and inappropriate behaviors), defined in the survey. These different topics were addressed over the general audience and within five specific categories of groups that are traditionally under-represented in astronomy: 1) job position (in particular early-career researchers), 2) expatriate/non-expatriate, 3) gender, 4) the belonging (or not) to another under-represented group, 5) parenthood.

This paper follows these themes and starts with general demographics in Section \ref{sec:Demographics} and then focuses on visibility and exposure in Section \ref{sec:Visibility and exposure at conferences}, recognition through publications and collaborations for projects in Section \ref{sec:Recognition with publications and projects}, disrespect in Section \ref{sec:Disrespect}, inappropriate behaviors in Section \ref{sec:Inappropriate behaviors}, and allyship in Section \ref{sec:Allies}. Eventually, the main conclusions of this study are summarized in Sec. \ref{sec:Conclusions} with some recommendations, issued both from the previous sections and from the comments sent by the survey respondents.

\section{Survey participant demographics}
\label{sec:Demographics}

The first questions of the survey (questions 1 to 7 in the Appendix \ref{sec:AppendixA}) aimed to draw a picture of the group of participants. The questions addressed respondents' professional status and topics (position and expertise), some geographic information (job location and expatriation), under-representation (or not) in the field (in terms of gender, but also other under-representation like ethnicity, disability, or sexual orientation), and parenthood. In designing the survey, we considered providing a precise definition and scope for the term `under-representation' (for instance, a question about ethnicity), since it would provide information about a specific discrimination factor. However, ultimately no formal definition was included for three reasons: 1) this question is prohibited in some countries like France (\href{https://www.lemonde.fr/les-decodeurs/article/2019/03/19/la-difficile-utilisation-des-statistiques-ethniques-en-france_5438453_4355770.html}{read this ``Le Monde" article}), 2) discriminating such a small community would release the anonymity of participants from under-represented groups by providing too many details (ex: female postdoctorate researcher of color), and 3) the outcomes would follow small number statistics and could not be significant at larger scales. Furthermore, having a statistic giving the percentage of people of color, LGBTQ+ people, and people with disabilities in the field is still a significant result. 

\begin{figure*}
   \begin{center}
   \begin{tabular}{cc}
   \includegraphics[width=8cm]{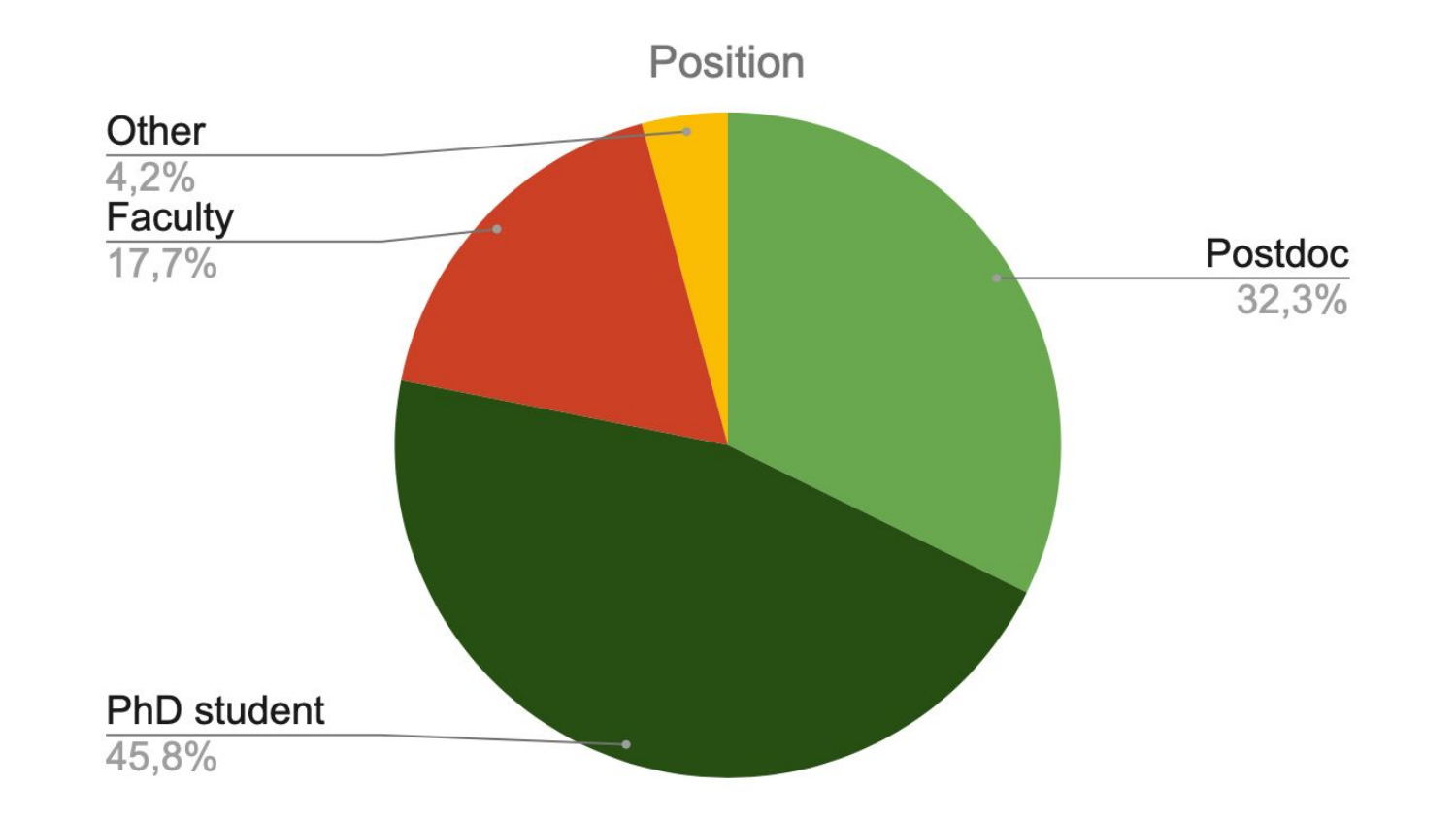} & \includegraphics[width=8cm]{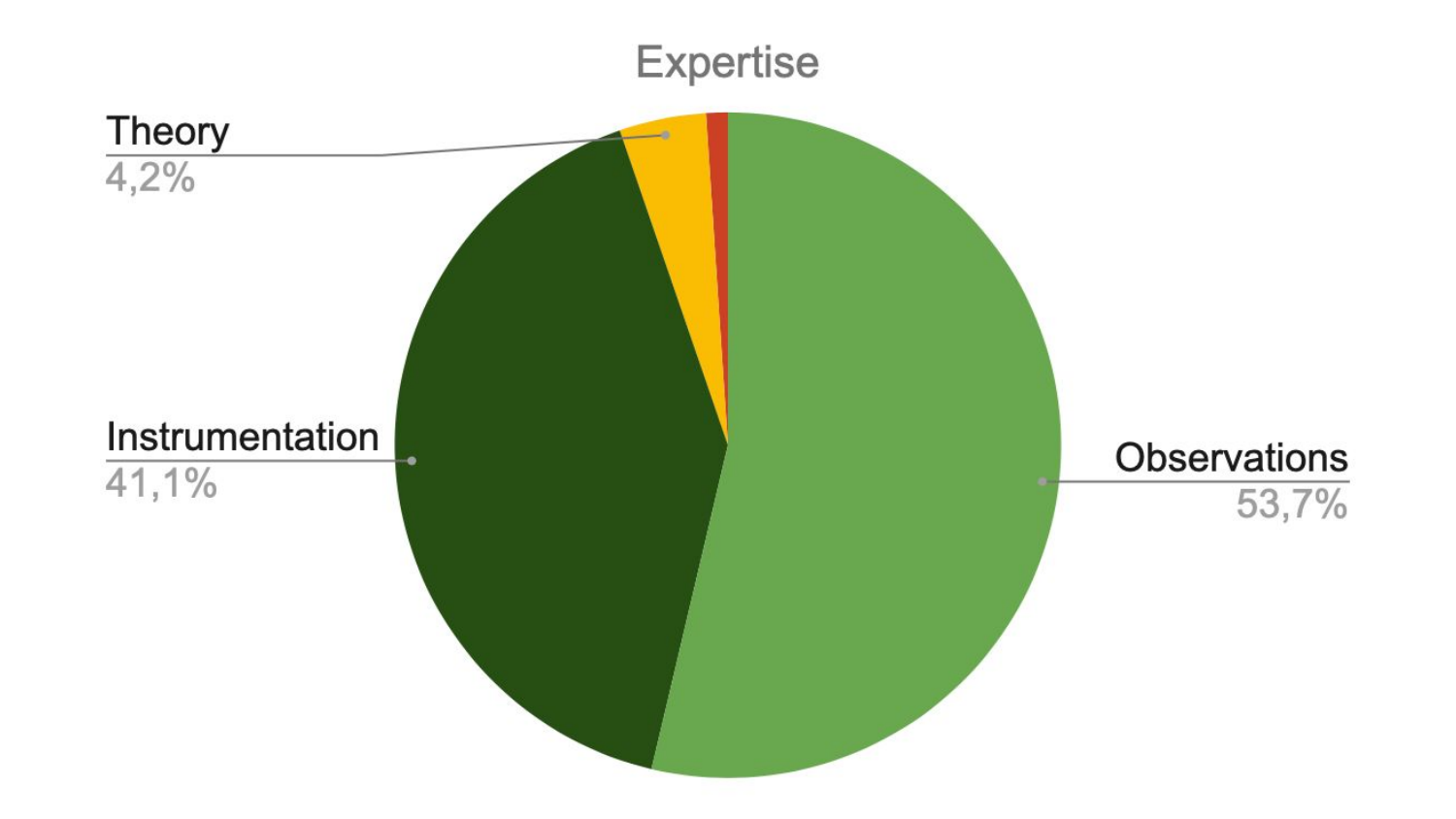} \\
   \includegraphics[width=8cm]{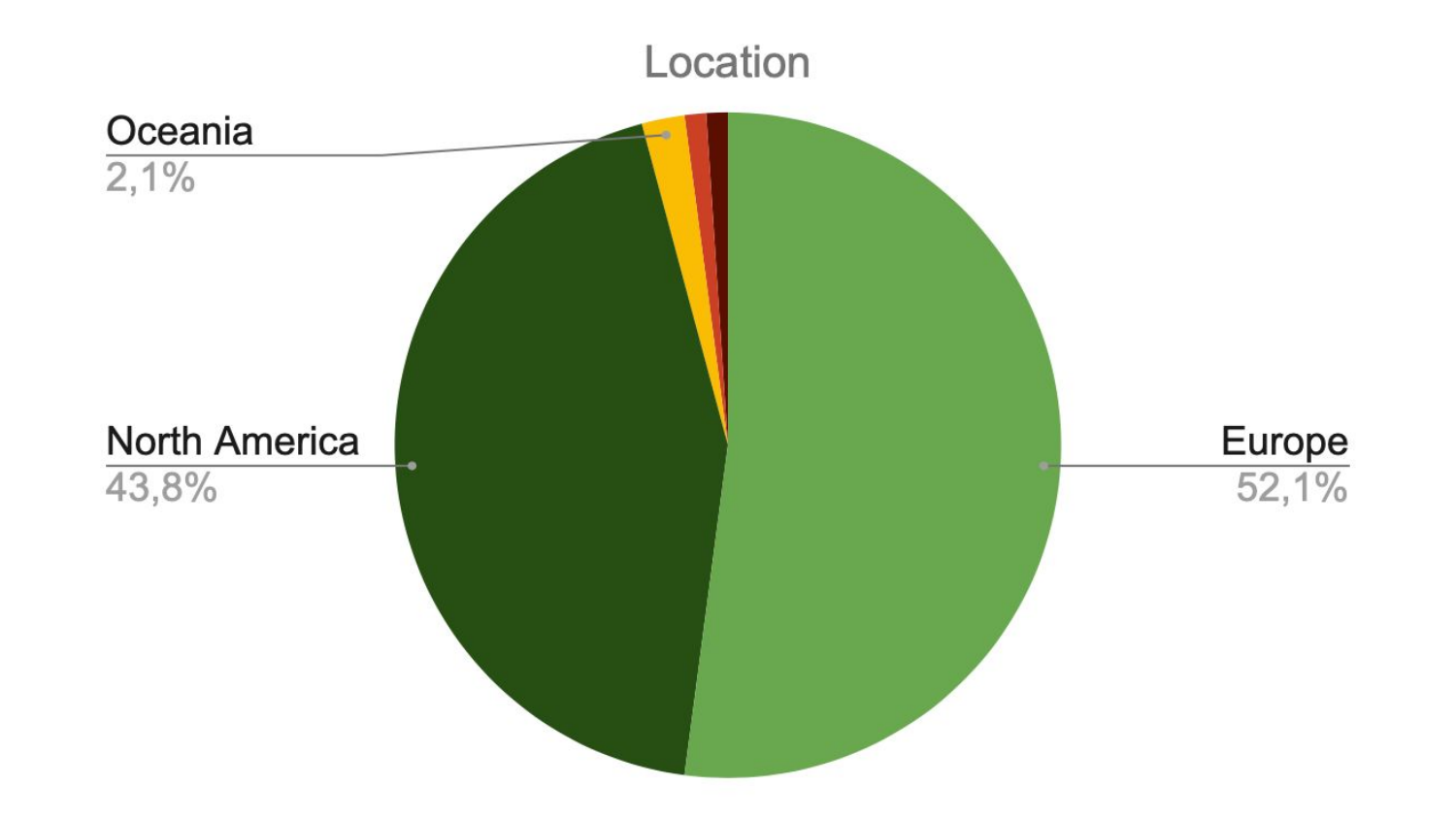} & \includegraphics[width=8cm]{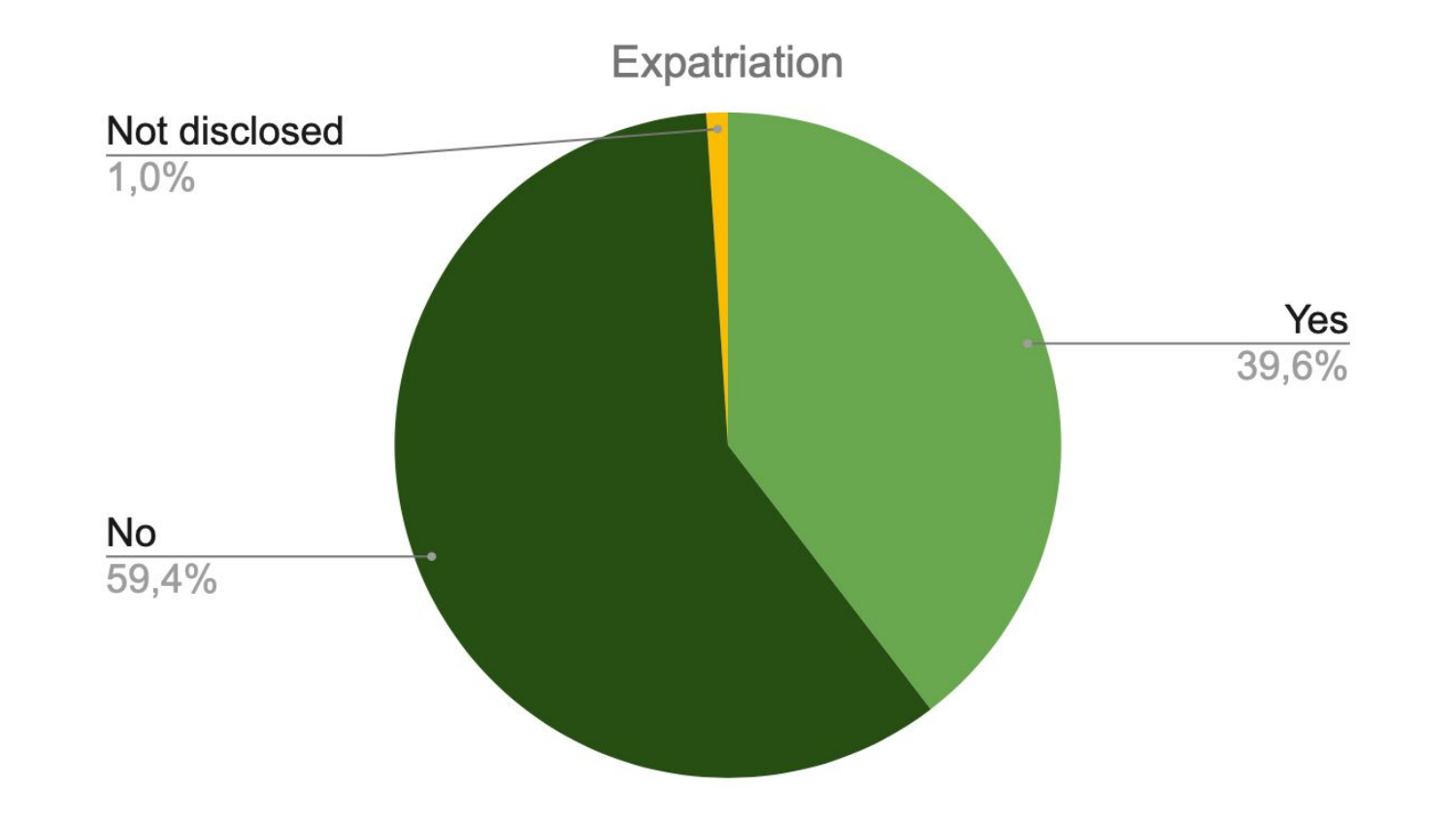} \\
   \includegraphics[width=8cm]{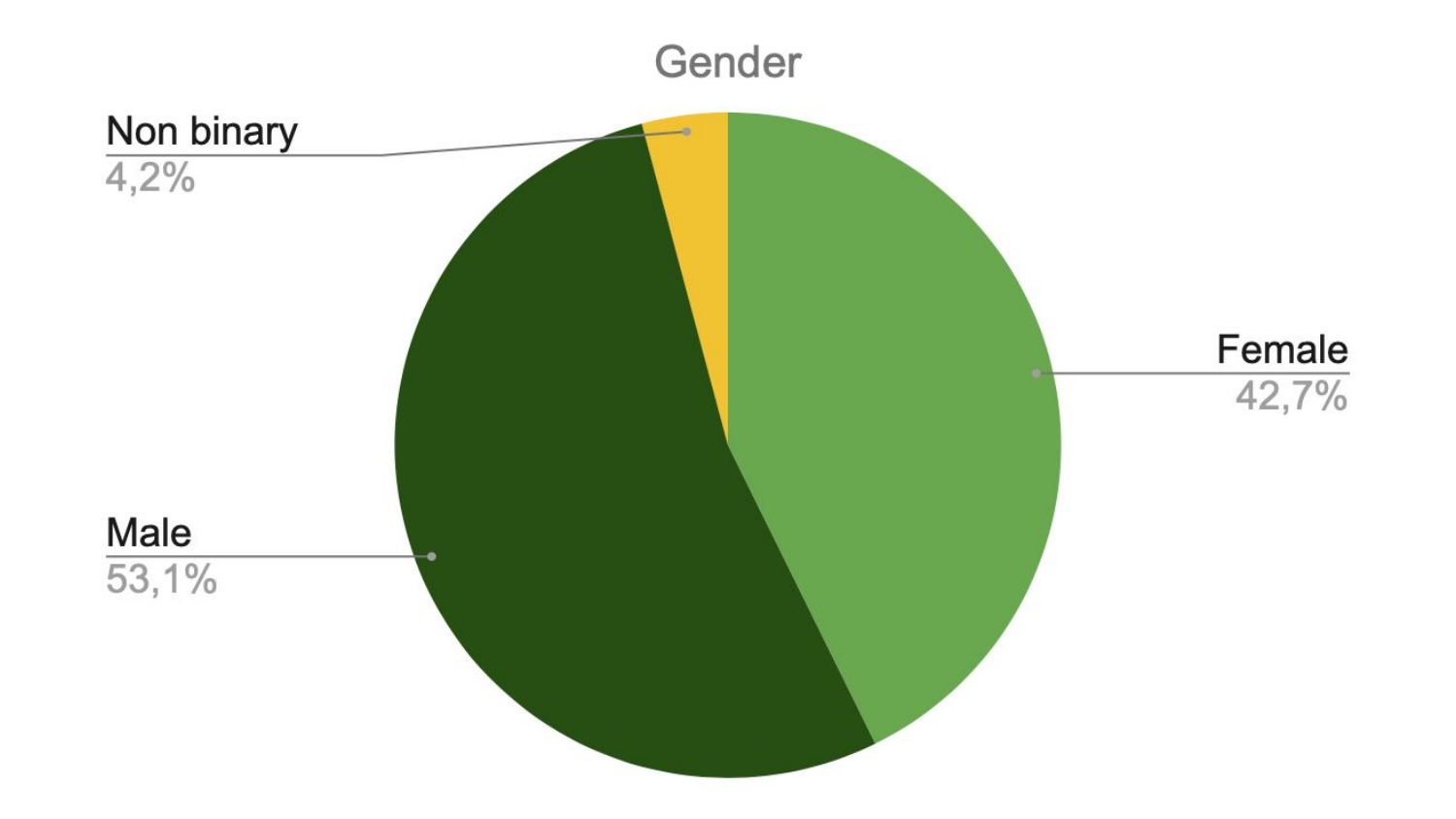} & \includegraphics[width=8cm]{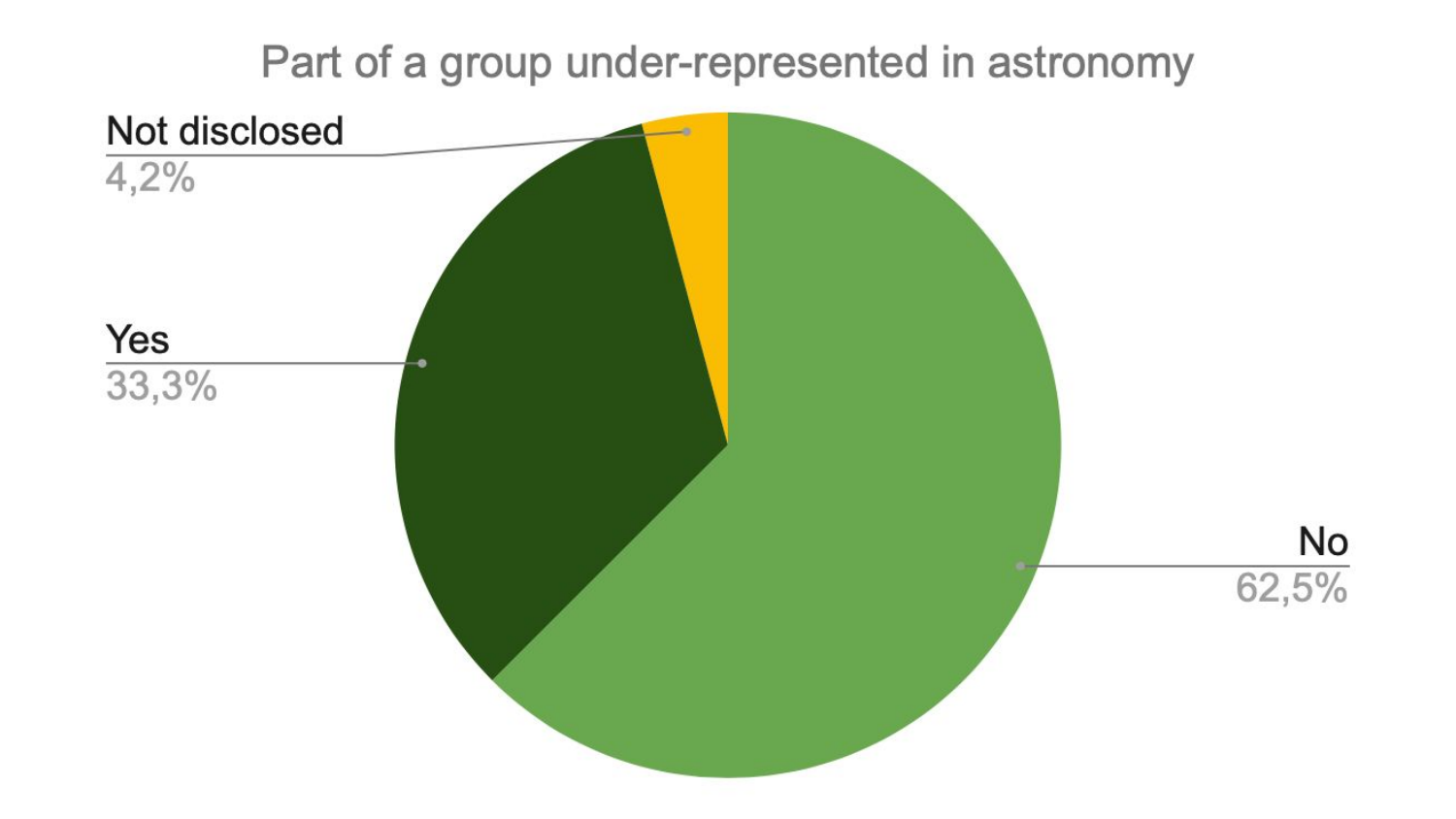} \\
   \end{tabular}
   \includegraphics[width=8cm]{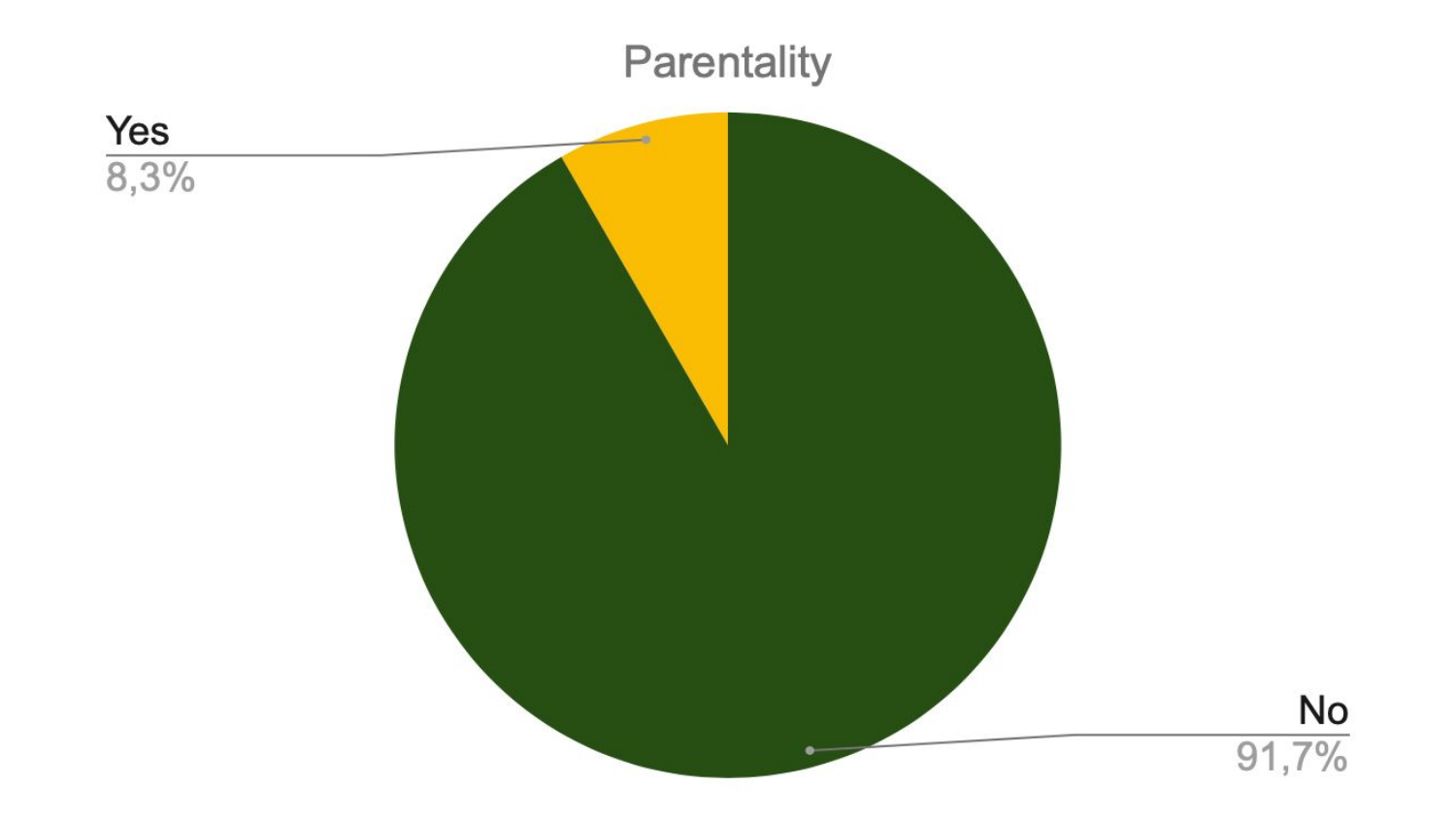}
   \end{center}
   \caption
   { \label{fig:Sec2_Fig1} 
Demographics of the survey participants, in terms of (top) position and expertise, (2nd line) geography, (3rd line) belonging to an under-represented community in astronomy and astrophysics, and (bottom) parenthood. Percentages which are not mentioned falls at $1\%$ (i.e. one participant).} 
   \end{figure*}

In total, this survey gathered $96$ responses among $218$ participants registered at the conference, yielding a participation rate of $44\%$. For comparison, the 2019 Spirit of Lyot conference survey gathered $100$ responses from $190$ participants (a participation rate of $53\%$). This slightly lower participation rate is not due to poor visibility of the 2022 survey: this new survey was mentioned and accessible during the registration process (while only during the conference week for the 2019 survey), and the 2019 results were presented over an invited talk at the 2022 conference with a direct incentive for filling this new 2022 survey. The demographic profiles of these $96$ participants are visible in Figure \ref{fig:Sec2_Fig1}:

$\bullet$ \textbf{Job position:} among respondents, $46\%$ are PhD students, $32\%$ postdoctorate researchers, $18\%$ faculty researchers, and $4\%$ of participants occupy other positions. Meanwhile, registrations at the conference indicate percentages around $37\%$ (PhD students), $26\%$ (postdoctorate researchers), $33\%$ (faculty researchers), and $4\%$ (other positions), which shows that the faculty group participated significantly less in the survey than the other groups. In the 2019 survey, faculty researchers had a higher participation rate, up to $47\%$.

$\bullet$ \textbf{Expertise:} Observations and Instrumentation are the main expertises of participants ($54\%$ and $41\%$ respectively), with Theory representing only $4\%$ of the participant research topics. One person did not answer the expertise question. In 2019, the Spirit of Lyot conference had more respondents in Instrumentation ($50\%$) than Observation ($46\%$), and an equivalent percentage of people working in Theory.

$\bullet$ \textbf{Location:} In a large majority of cases, participants work and live in Europe ($52\%$) and North America ($44\%$), followed by Oceania ($2\%$), Asia ($1\%$, so one person), and South America (also $1\%$). Since the conference was hosted in the Netherlands, large representation from European institutes is not surprising and in general, the number of participants from Asia and Oceania is very low compared to Europe and North America. Similarly, the 2019 conference survey had more participants from Asia ($8\%$), and the conference was hosted in Japan.

$\bullet$ \textbf{Expatriation:} $40\%$ of participants indicate being expatriates, versus $59\%$ indicate not being expatriates. 

$\bullet$ \textbf{Gender:} Among the participants, $43\%$ are female (F), $53\%$ are male (M), and $4\%$ are non-binary (NB). This data does not adequately represent the general ratio of women in the field. For instance, women are $21\%$ of International Astronomical Union (IAU) members (data from January 2023, \href{https://www.iau.org/public/themes/member_statistics/#4}{access to the IAU Member Statistics}). From the conference registration data (participants' self-reported pronouns), we estimate the conference attendance to be around $32\%$ female, $67\%$ male, and $1\%$ non-binary, which indicates that male participants under-participated in the survey. The 2019 statistics indicated a better representation of the conference audience with $29\%$ of responses from women, $69\%$ from men, and $2\%$ from non-binary people.

$\bullet$ \textbf{Under-representation:} In the survey, under-representation was defined as considering oneself as part of a group under-represented in astronomy in terms of ethnicity, disability, or sexual orientation specifically (no gender, expatriation, etc.). $33\%$ of participants answered that they are part of an under-represented community in astronomy versus $63\%$ answered that they are not. 

$\bullet$ \textbf{Parenthood:} $8\%$ of participants are parents, $92\%$ are not. 

Later in this paper, we focus on specific demographic data: job position, expatriation status, gender, under-representation, and parenthood. For the gender criterion, including non-binary people into survey outcomes has been proven to be crucial, particularly since they experience specific biases in the field \citep{Rasmussen2019, Strauss2020}. However, the nature of this paper, which focuses on discrimination and inappropriate behaviors, raises concerns about releasing anonymity due to small number statistics, given that $4$ participants identify as non-binary. To ensure their anonymity, we combine the data of both under-represented gender groups (female and non-binary) later on in this paper. We nevertheless acknowledge the specificity and seriousness of biases happening against non-binary people, and the need for data large enough to address the specific issues they encounter. For the same reasons, the data from the people occupying `other' jobs than PhD, postdoc, or faculty, have not been used since there are $4$ people in this category and there was ambiguity in grouping them with another demographic: they could be permanent engineers or under-graduate students, for instance.

\begin{figure}
   \begin{center}
   \includegraphics[width=8.5cm]{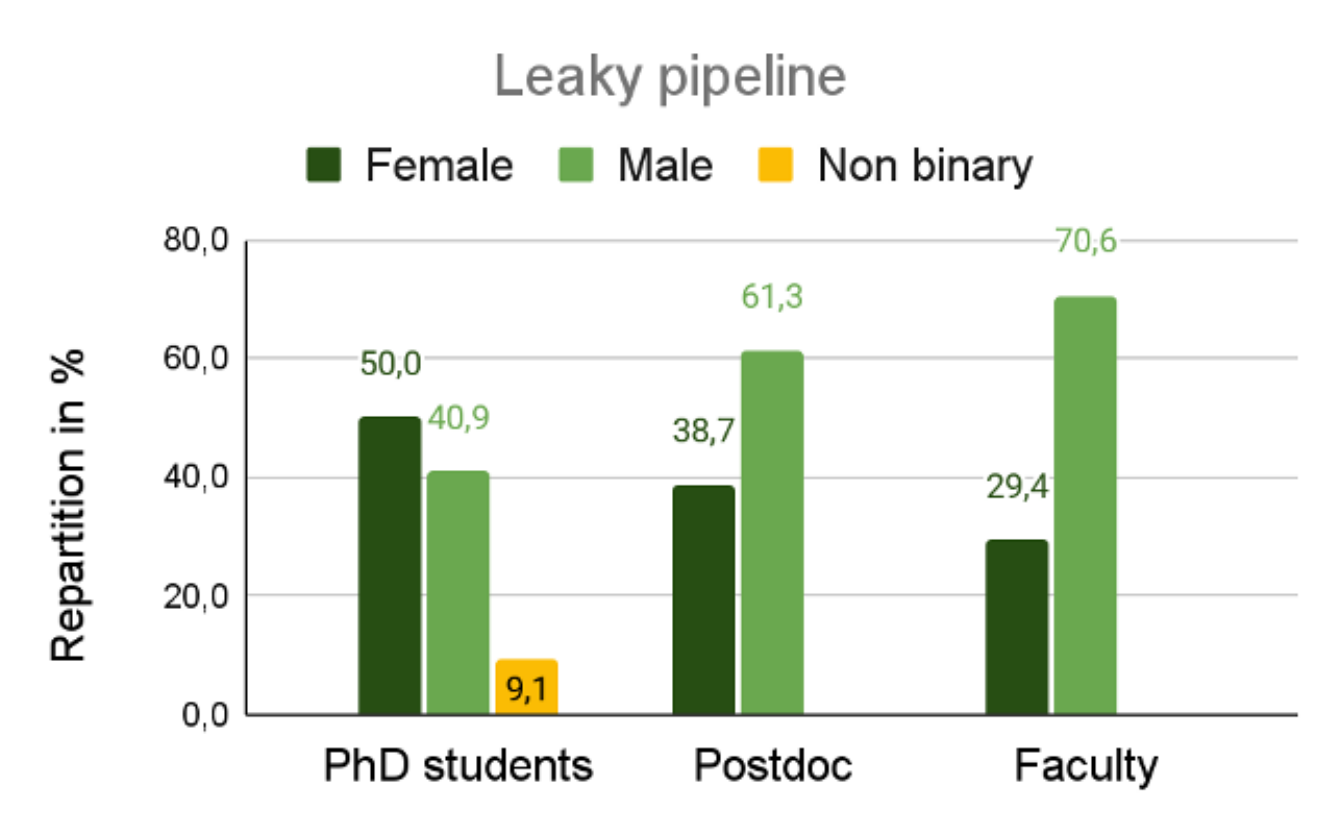}
   \end{center}
   \caption
   { \label{fig:Sec2_Fig3} 
The `leaky pipeline', demonstrating the drop-out rates as a function of increasing career stage, is illustrated with data from the current survey. In addition to showing this real phenomenon in astronomy, it indicates a correlation between job position and gender in the interpretation of the outcomes of the survey later in this paper.} 
   \end{figure}

Some of these categories are not fully independent from each other and correlations should be pointed out, which can make the data analysis more complicated later on. For instance, among the $8$ parents, $7$ are male and only $1$ is female, so parenthood and gender can impact each other's outcomes. Similarly, $5$ of the $8$ parents are faculty researchers, so parents tend to occupy permanent positions and be more senior. Another example: expatriates tend to be postdocs more often than other job categories ($65\%$ of postdocs are expatriates versus $40\%$ over the whole respondent panel). The last example from this survey is based on the correlation between gender and position, well-known and studied as the ``leaky pipeline" \citep{DOrgeville2014}: Figure \ref{fig:Sec2_Fig3} illustrates this effect with data from the current survey. The percentage of women and non-binary people is decreasing with career step, from $50\%$ (female) and $9\%$ (non-binary) at the PhD stage to $29\%$ (female) and $0\%$ (non-binary) at the permanent position (for this figure, the work position `other' has been removed). The phenomenon is complex to study, since it can be due to both women and non-binary people leaving the field more than men, and/or to a recent increase of women at the beginning of their careers in astronomy. In 2019, this was also measured, and monitoring a few years later was proposed in order to evaluate if women and non-binary PhD students had stayed in the field; according to this new data, the lack of women at higher career steps was not even slightly compensated for, and a longer term follow-up is necessary. These various dependencies should be kept in mind later when analyzing the data per category.

Another potential bias should be considered in the processing and analysis of the data: while most questions refer to a defined time period of $2.5$ years (from January 2020 to June 2022), some respondents may have entered the field or changed positions during this timeframe. This could affect certain results, particularly those related to career stages. For example, there may be a bias for first- or second-year PhD students who have been in the field for less than $2.5$ years, or cases of data leakage between career stages — such as an event experienced by someone who was a postdoc at the time but is now recorded as a permanent researcher. By contrast, the 2019 survey covered respondents’ entire careers. The decision to define a more limited period for this new survey was made precisely to help reduce such biases.

A final compromise was made in the results presented below concerning parents: they should not be considered a homogeneous group, as parenthood appears to affect male and female parents differently \citep{Nicot2009}. Indeed, parenthood tends to negatively impact women's careers (the so-called "motherhood penalty"), while having a positive effect on men's careers (the "fatherhood bonus") \citep{Gangl2009}. As a result, in this study, we consider the experiences of mothers and fathers separately. However, due to the very small number of mothers in the sample (only one person), only the group of fathers (7 people) is included in the analysis. This decision is again based on the importance of both having a sufficiently large sample and preserving the anonymity of the mother. It should be noted here that the low number of mothers may reflect the difficulties they face in attending conferences.

\section{Visibility and exposure at conferences}
\label{sec:Visibility and exposure at conferences}

In this section, covering questions 8 to 11 of the survey (see Appendix \ref{sec:AppendixA}), we explore conference attendance, talks awarded, invited seminars, and invitations to join international conference Scientific Organizing Committees (SOC) as four major metrics for visibility within the field.

Table \ref{table:Sec3_Visibility} presents the outcomes of these metrics, with 1) the average number of conferences per person since 2020, 2) the average number of conference talks per person since 2020, 3) the average number of invitations to give seminars per person since 2020, and 4) the percentage of people within the category invited to join a conference SOC since 2020. The first line of the table indicates the metrics computed for all survey participants and can be used as a reference to compare the following line indicators, which are computed per category. 

A larger percentage of survey participants reported attending a conference in the years 2020-2022 than 2017-2019. Given the travel restrictions imposed by the COVID-19 pandemic, this may be surprising. However, the resulting possibilities for remote participation increased conference attendance rates and inclusion \citep{Foramitti2021}. While a greater equity in conference attendance does not equate to equal participation (e.g. disproportionate participation in Q\&A sessions across genders \citep{Jarvis2023}), this is an encouraging trend.

\begin{table*}
\centering
\begin{tabular}{|c|c|c|c|c|}
  \hline
   & \textbf{\#conf/person} & \textbf{\#talk/person} & \textbf{\#seminar/person} & \textbf{$\%$ invited in SOC} \\
  \hline
  \hline
  General & $2.8$ & $1.5$ & $1.4$ & $12.1$ \\
  \hline
  \hline
  PhD student & $2.4$ & $0.9$ & $0.5$ & $0.0$ \\
  Postdoc & $3.3$ & $2.3$ & $2.2$ & $3.2$ \\
  Faculty & $2.9$ & $1.8$ & $2.2$ & $52.9$ \\
  \hline
  \hline
  Expatriate & $3.0$ & $1.7$ & $1.4$ & $8.1$ \\
  Non-expatriate & $2.5$ & $1.4$ & $1.3$ & $13.2$ \\
  \hline
  \hline
  Female + non-binary & $3.1$ & $1.4$ & $1.5$ & $11.9$ \\
  Male & $2.5$ & $1.7$ & $1.3$ & $12.2$ \\
  \hline
  \hline
  Under-represented & $2.8$ & $1.3$ & $1.4$ & $20.0$ \\
  Not under-represented & $2.7$ & $1.5$ & $1.3$ & $8.8$ \\
  \hline
  \hline
  Father & $2.0$ & $2.2$ & $2.1$ & $28.6$ \\
  Non parent & $2.8$ & $1.5$ & $1.3$ & $9.6$ \\
  \hline
\end{tabular}
\caption{Survey results about conference-related visibility: access to conference (average number of conferences since 2020 per person of the category), recognition of one's expertise through an oral contribution (average number of conference talks since 2020 per person of the category), invitation to give seminars (average number of seminar since 2020 per person of the category), and percentage of people from the category invited to join a conference SOC since 2020.}
\label{table:Sec3_Visibility}
\end{table*}

$\bullet$ \textbf{Results per career stage:} Overall, within $2.5$ years (January 2020 to June 2022), all position participants had access to conferences. Non-permanent researchers, in particular postdocs, attended several conferences, in spite of conference and/or travel cancellations due to COVID-19. One can still notice that PhD students tend to have less talks than the other categories, in contrast to postdoctorate and faculty researchers. It is worth noting that there may be a bias, as first-year PhD students often do not have the opportunity to give a talk, given that they may still be in the process of obtaining their first research results. For postdocs, this result is in line with the 2019 survey, which indicated a healthy community that encourages and supports postdocs in accessing conferences. This is necessary since postdocs are required to actively advertise their work and expand their network in search of a new or permanent position. In 2019, the survey concluded that PhD students are the least promoted through conference attendance. Even though numbers for both conferences cannot be compared, we should remain conscious about exposure of PhD students at conferences, since they also need to promote their work and find collaborations and job opportunities. In addition, SOC invitations almost only include permanent researchers: although being in an SOC requires professional experience and a network, early-career scientists could be included with their own network, experience, and perspective.

$\bullet$ \textbf{Results per expatriate status:} Expatriate people are more visible at conferences (attendance and talks), which could be due to the correlation of this category with the postdoc one. However, mostly non-expatriate people are invited to join an SOC, (only $13.2\%$ of non-expatriate respondents say they were invited to join an SOC). The community would benefit from more inclusion in SOCs.

$\bullet$ \textbf{Results per gender:} There is clearly higher attendance to conferences by women and non-binary people ($3.1$ conferences/respondent) than by men ($2.5$), however this higher participation does not impact the average number of talks nor invited seminars per female or non-binary person. For SOC invitations, no clear trend emerges here, in opposition with the 2019 survey which showed that women were more solicited in SOC: $21\%$ of the female participants versus $14\%$ of the male participants were invited to an SOC in 2018.

$\bullet$ \textbf{Results according to community representation:} Access to conferences, talks, and seminars appears quite equivalent for participants from represented and under-represented groups. However, one can notice an over-solicitation to join SOCs for people from under-represented groups despite the fact that they are slightly less likely to get a talk at conferences. It is a good sign that people from under-represented groups are asked to join a SOC. However, it can also lead to an over solicitation of these individuals, burdening them with non-research-related tasks. This also shows that even a diverse panel can exhibit biases when selecting speakers.

$\bullet$ \textbf{Results for fathers vs non parents:} Fathers appear to be more selective than non-parents, meaning they tend to go to conferences where they will give a talk. They are also largely invited to join SOCs, which could be an effect of their permanent status ($4$ out of $7$ fathers are faculty researchers).

\textbf{$\rightarrow$ Recommendation 1: Being more inclusive by opening the organization of conferences to expatriate and non-permanent researchers.}

\textbf{$\rightarrow$ Recommendation 2: Parents should be supported in attending conferences through the provision of childcare resources. This support would be especially impactful for mothers, who are currently underrepresented (1 mother, 7 fathers).}

\section{Recognition with publications and projects}
\label{sec:Recognition with publications and projects}

This section corresponds to questions 12 to 16 of the survey (see Appendix \ref{sec:AppendixA}). The objective is to evaluate the publication rate and the recognition of people's expertise through different criteria: the number of requests to review papers, their feeling of being unfairly absent or present in co-author lists, and their feeling of being left out from publications and projects, all since January 2020 (within a range of $2.5$ years).

\begin{figure*}
   \begin{center}
   \begin{tabular}{ccc}
   \includegraphics[width=4.9cm]{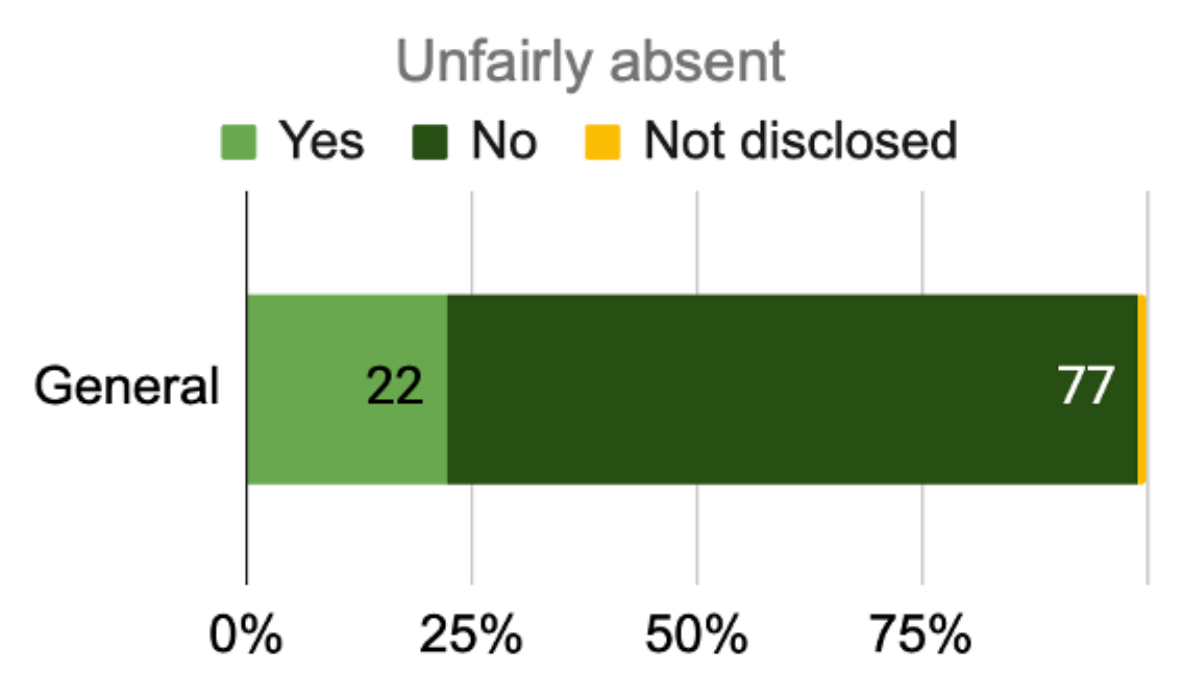} & \includegraphics[width=4.9cm]{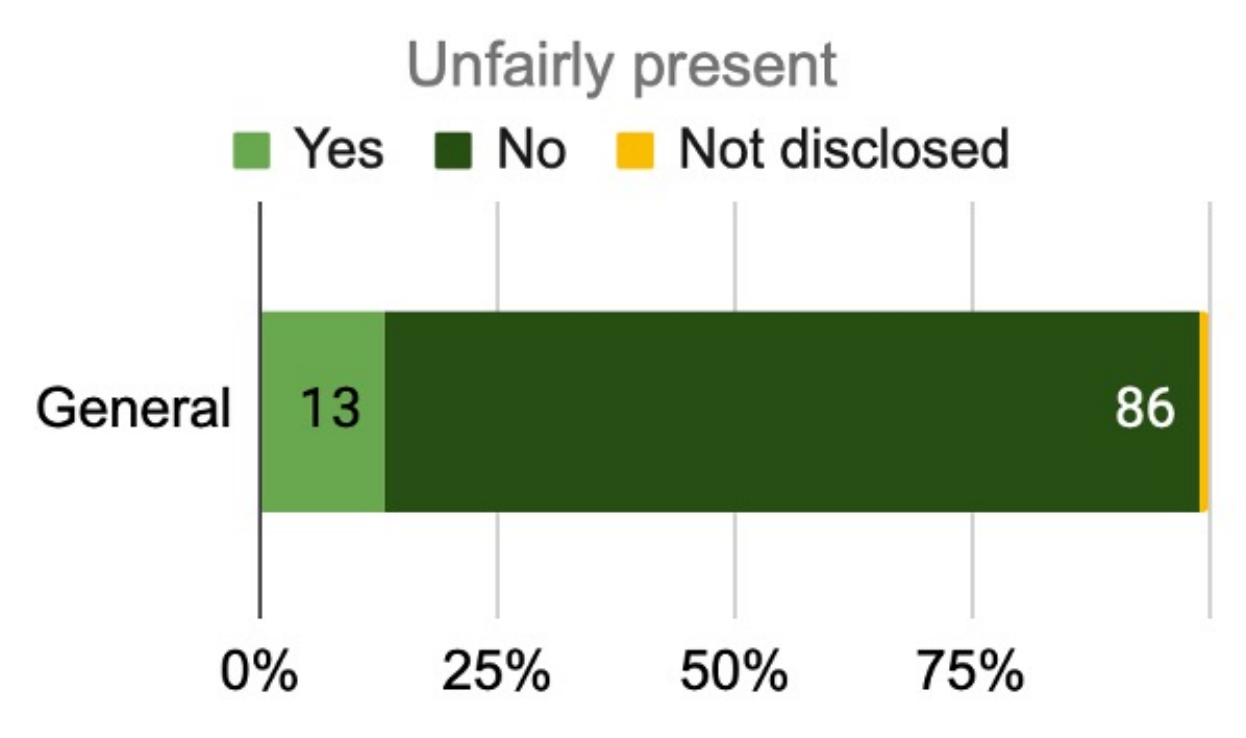} & \includegraphics[width=5.2cm]{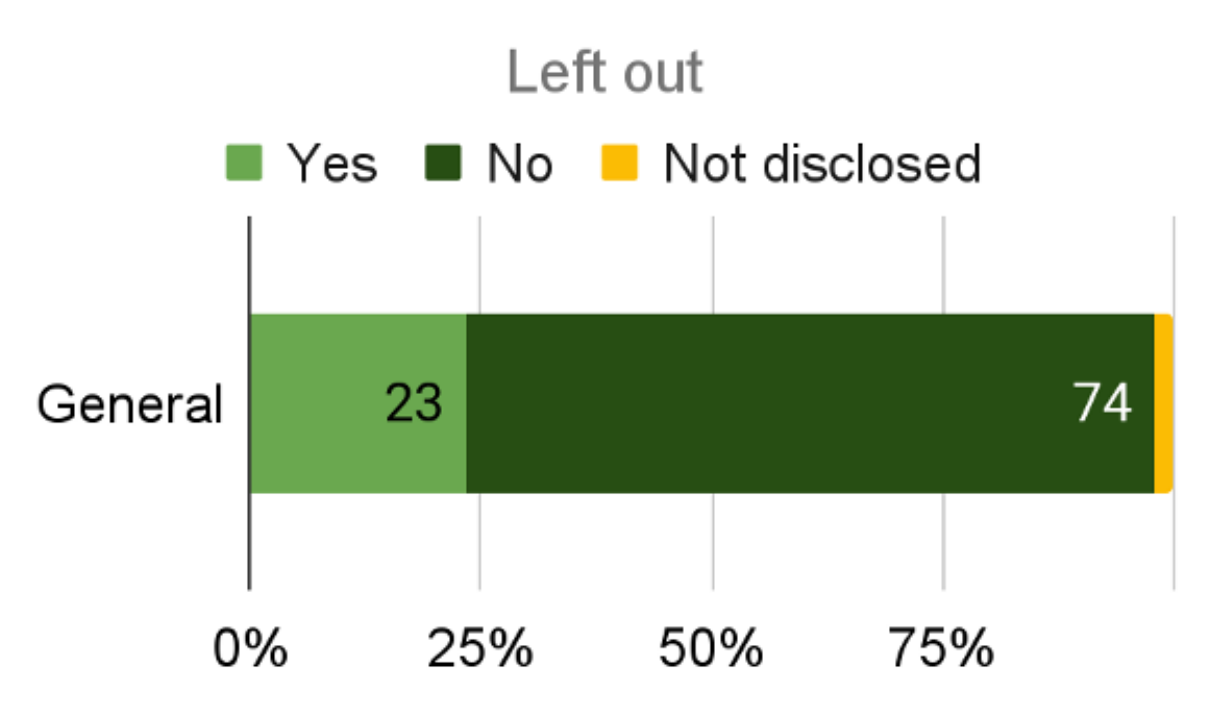} \\
   \includegraphics[width=5.2cm]{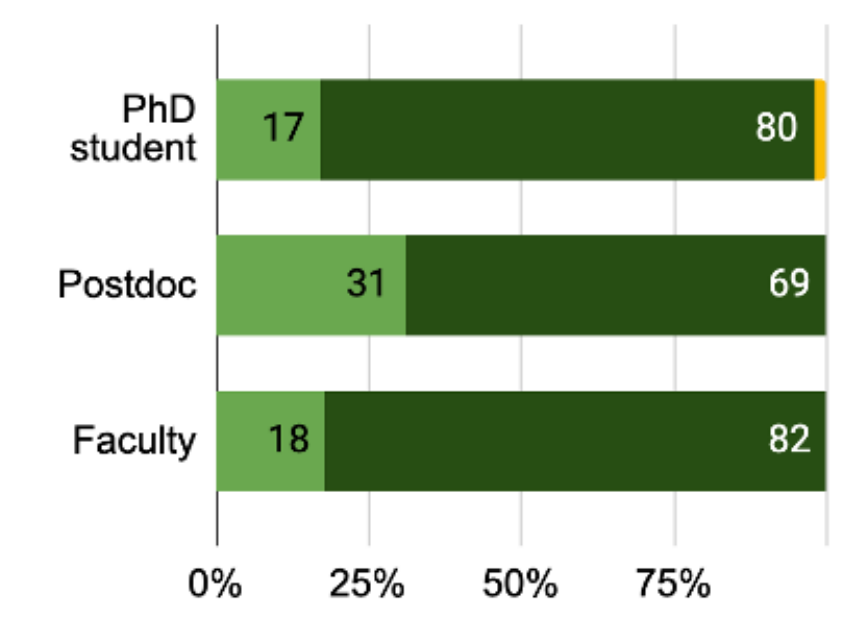} & \includegraphics[width=5.2cm]{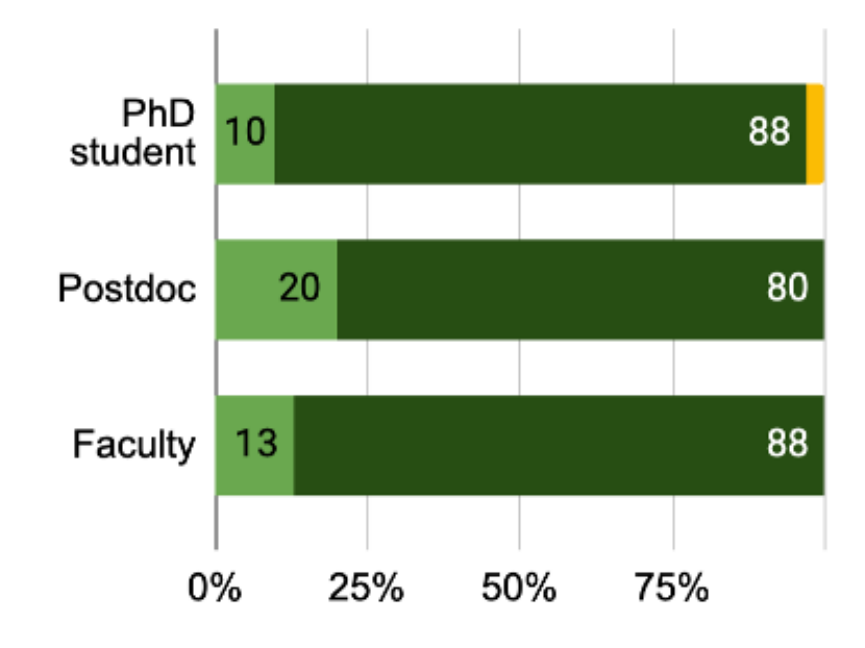} & \includegraphics[width=5.2cm]{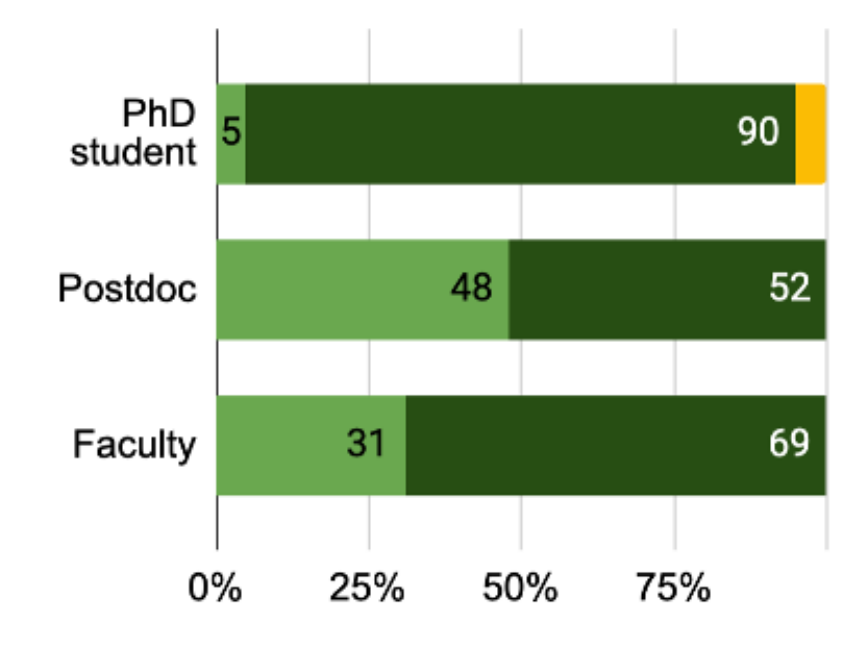} \\
   \includegraphics[width=5.2cm]{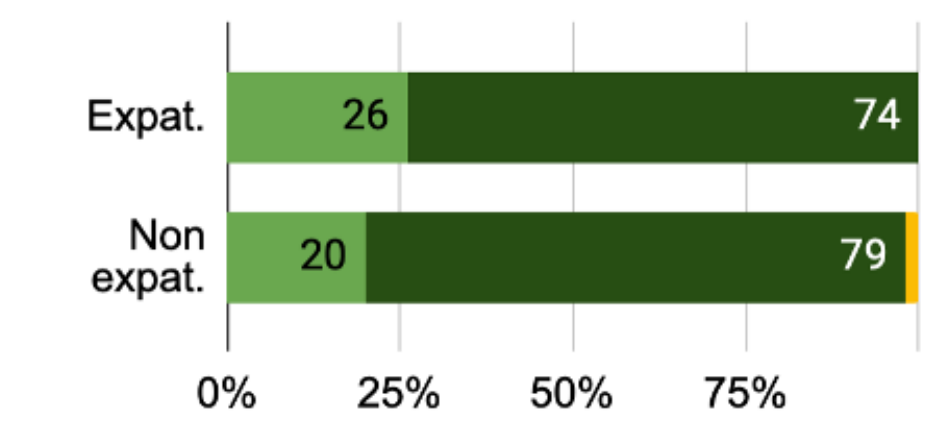} & \includegraphics[width=5.2cm]{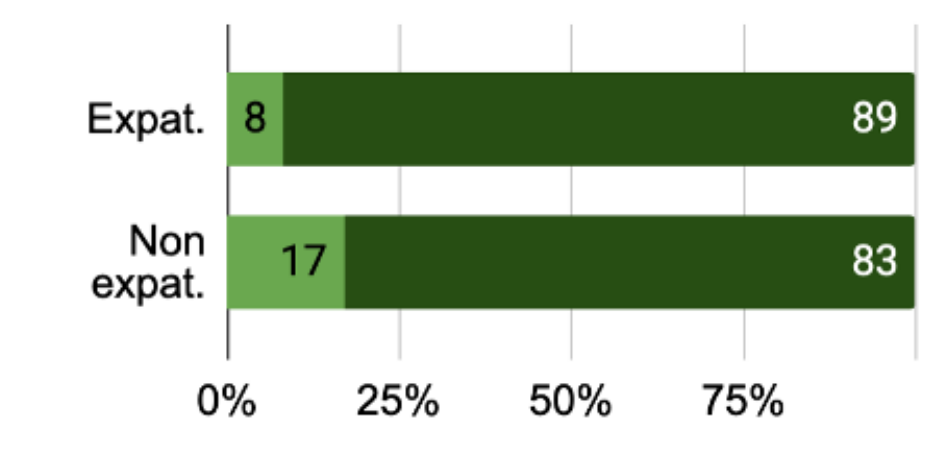} & \includegraphics[width=5.2cm]{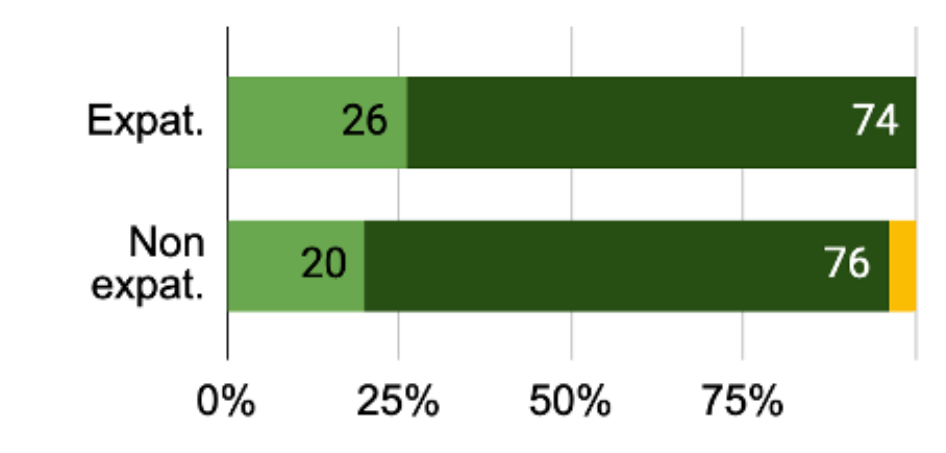} \\
   \includegraphics[width=5.2cm]{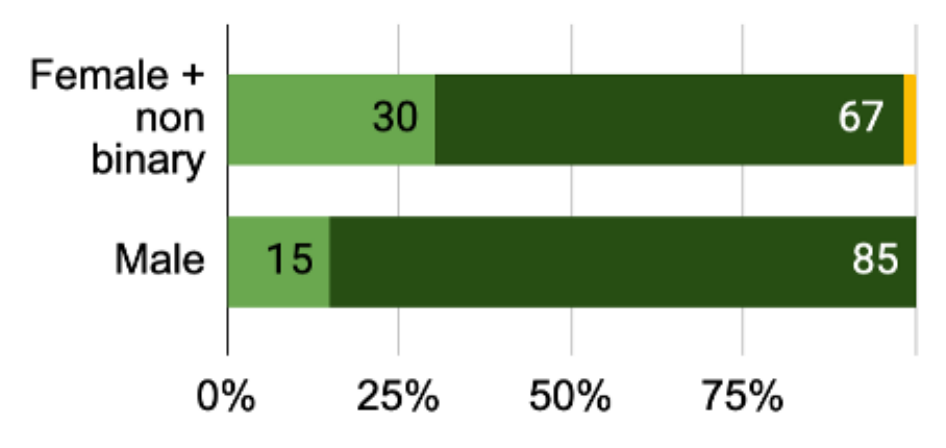} & \includegraphics[width=5.2cm]{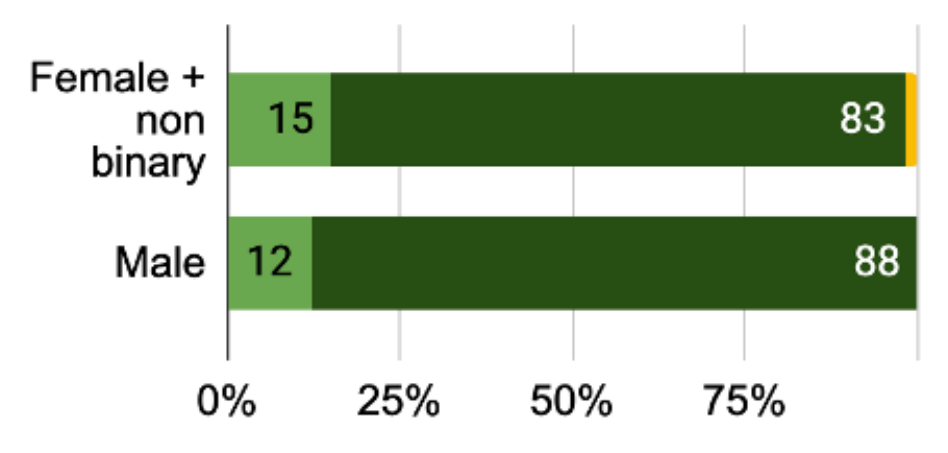} & \includegraphics[width=5.2cm]{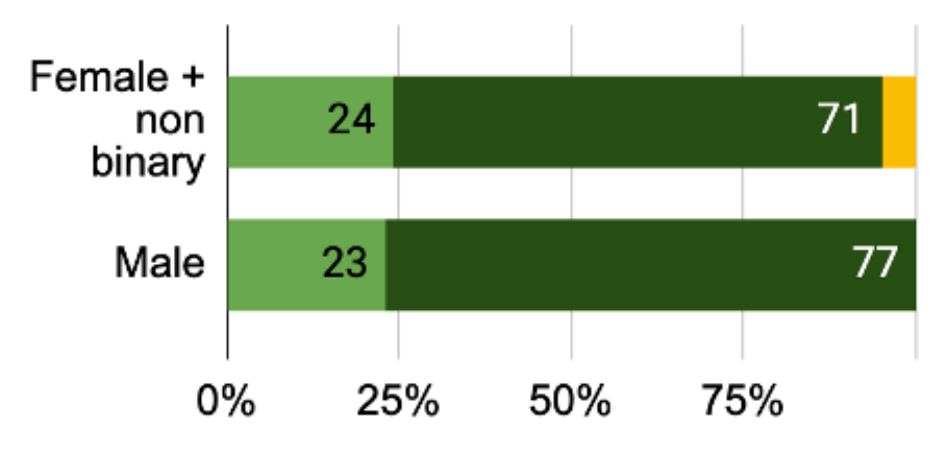} \\
   \includegraphics[width=5.2cm]{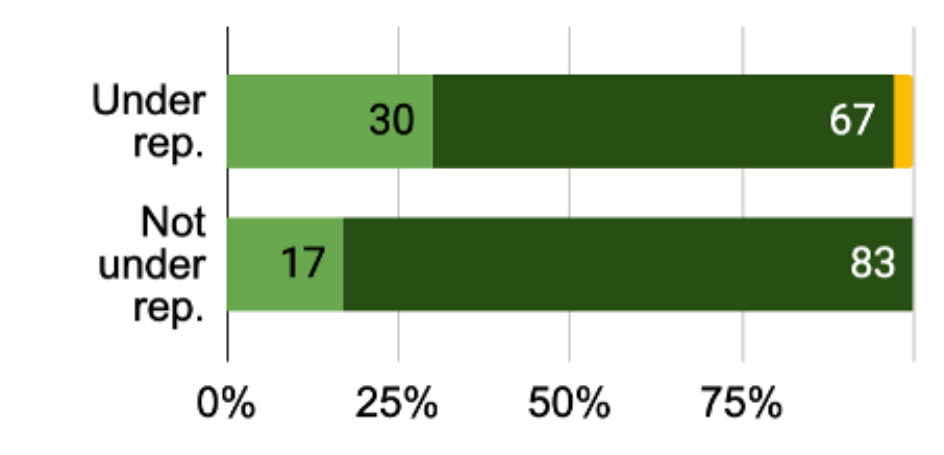} & \includegraphics[width=5.2cm]{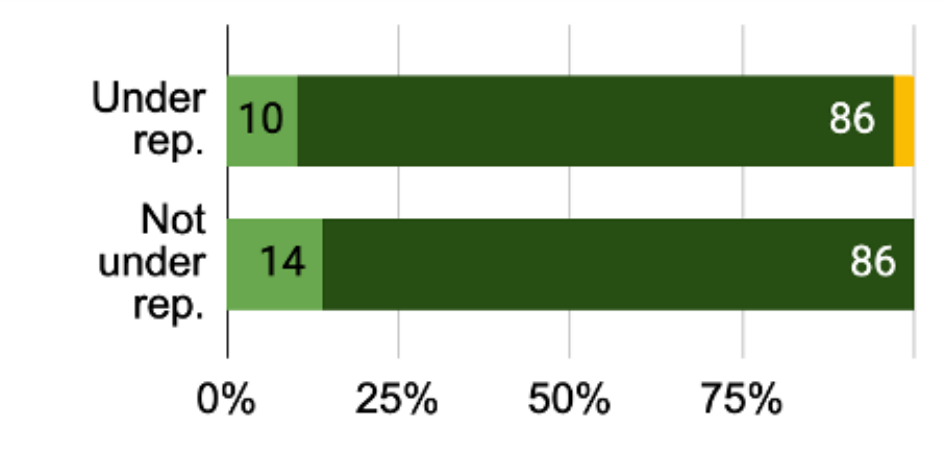} & \includegraphics[width=5.2cm]{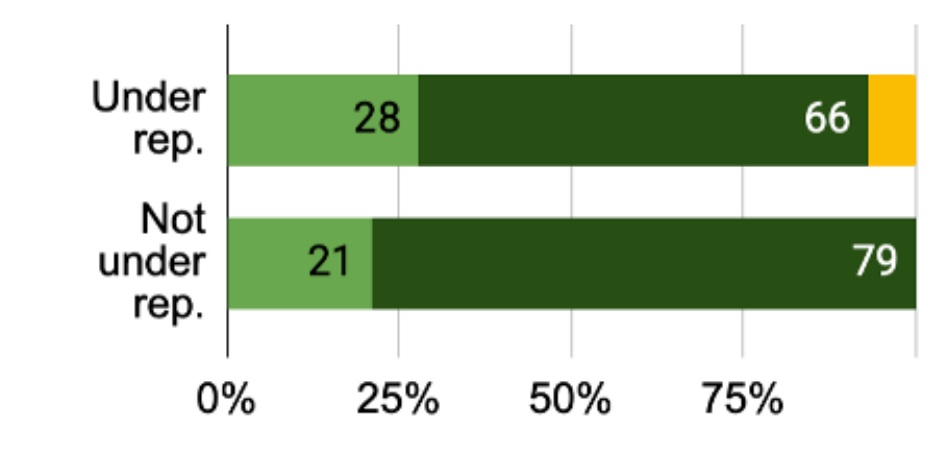} \\
   \includegraphics[width=5.2cm]{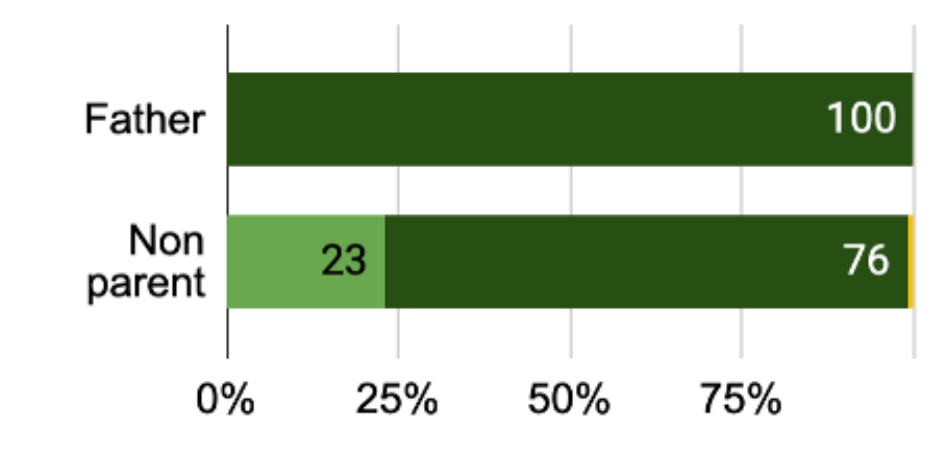} & \includegraphics[width=5.2cm]{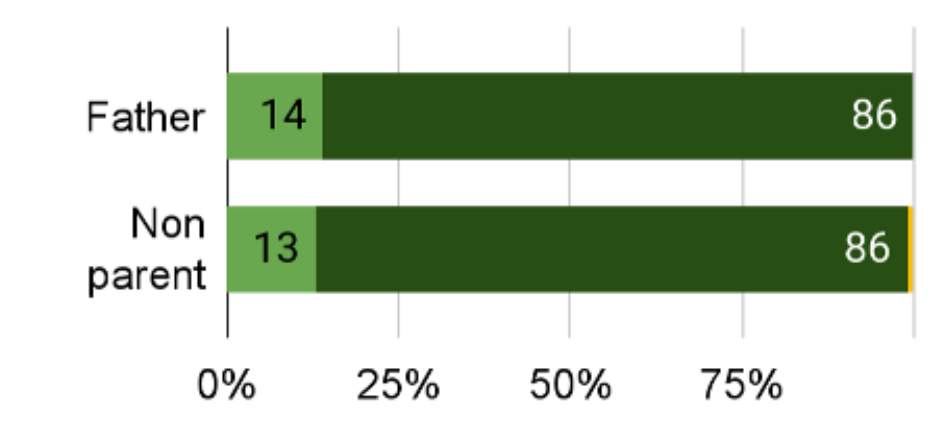} & \includegraphics[width=5.2cm]{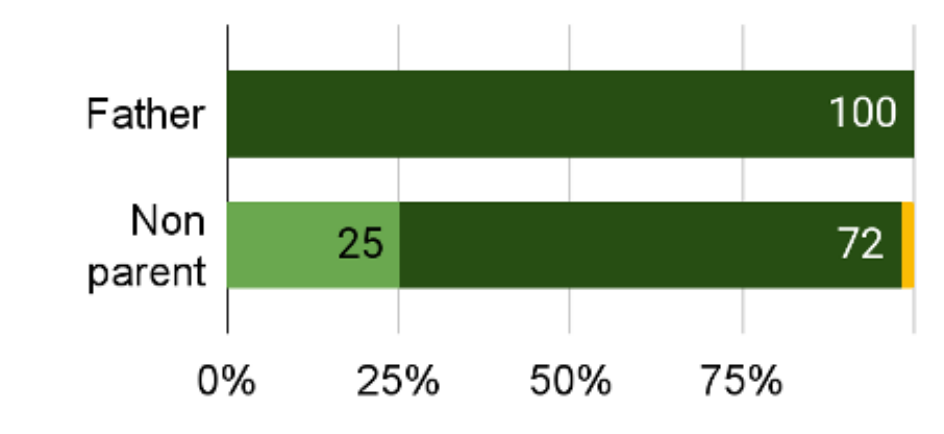} \\
   \end{tabular}
   \end{center}
   \caption 
   { \label{fig:Sec4_Fig1} 
Expertise recognition through inclusion in publications and projects, since January 2020: (left) percentage of people, per category, having felt unfairly absent from a co-author list, (center) percentages of people, per category, having felt unfairly present in a co-author list, (right) percentages of people, per category, having felt left out from a publication or project.} 
   \end{figure*}

The outcomes are presented in Table \ref{table:Sec4_Recognition} and Figure \ref{fig:Sec4_Fig1}. Table \ref{table:Sec4_Recognition} presents the average number of peer-reviewed papers published since January 2020 per person as first author and the average number of reviewing requests since January 2020, for the entire participant group (top line) and within specific groups (other lines). Figure \ref{fig:Sec4_Fig1} presents the percentages of people having felt unfairly absent (left column) and present (center) in co-author lists and having felt left out from a publication or project (right), all since January 2020.

From the entire participant group (top line of both the table and the figure), a notable number is that people publish on average one peer-reviewed paper within $2.5$ years. One can also notice that almost one quarter of participants have felt excluded from co-authorship, and one quarter from publications, or projects. Despite remaining very problematic, this is lower than in the 2019 survey (30\% of respondents had felt unfairly excluded from author lists), even if both questions did not cover the same time range (no time range was specified in 2019).

\begin{table*}
\centering
\begin{tabular}{|c|c|c|}
  \hline
   & \textbf{\#articles/person} & \textbf{\#review requests/person} \\
  \hline
  \hline
  General & $1.0$ & $1.4$ \\
  \hline
  \hline
  PhD student & $0.6$ & $0.03$ \\
  Postdoc & $1.7$ & $2.2$ \\
  Faculty & $0.8$ & $3.5$ \\
  \hline
  \hline
  Expatriate & $1.3$ & $1.9$ \\
  Non-expatriate & $0.8$ & $1.1$ \\
  \hline
  \hline
  Female + non-binary & $1.0$ & $0.9$ \\
  Male & $1.0$ & $1.9$ \\
  \hline
  \hline
  Under-represented & $0.7$ & $1.1$ \\
  Not under-represented & $1.0$ & $1.5$ \\
  \hline
  \hline
  Father & $1.6$ & $3.3$ \\
  Non parent & $0.9$ & $1.3$ \\
  \hline
\end{tabular}
\caption{Recognition in terms of publications and reviewing: publication rate (average number of peer-reviewed articles since 2020 per person of the category) and recognition of one's expertise through reviewing requests (average number of reviewing requests since 2020 per person of the category).}
\label{table:Sec4_Recognition}
\end{table*}

$\bullet$ \textbf{Results per career stage:} Researchers from almost all categories have felt biased against, and even more specifically postdoctorate researchers, whose percentages of feeling excluded are quite dramatic: one third of postdocs have felt unfairly absent from co-authorships since January 2020, and half of them have felt left out from publications or projects. We notice that even one third of permanent researchers feel excluded from publications or projects. This trend is similar, even if less extreme, to that in the 2019 survey (which did not specify a restricted time period). We should therefore be particularly careful about the inclusion of non-permanent researchers in publications, since their access to positions and grants depends on their curriculum and, among other elements, on their capacity to promote their work through publications and to join projects and collaborations.

$\bullet$ \textbf{Results per expatriate status:} Expatriate people are slightly more impacted by exclusion from co-authorship, publications, and projects than non-expatriates.

$\bullet$ \textbf{Results per gender:} Female and non-binary people feel excluded from co-authorships twice as often as their male colleagues. This result is quite new, since the 2019 survey did not show a significant difference in inclusion in author lists between genders. Men are also twice as requested to review papers as female and non-binary people.

$\bullet$ \textbf{Results according to community representation:} Once again, a clear trend appears here, since people from under-represented groups feel unfairly absent from co-authorships almost twice as often as their colleagues.

$\bullet$ \textbf{Results for fathers vs non parents:} Fathers feel less isolated from co-author lists, publications, and projects than non-parents, with a possible correlation with their job position. With only $7$ fathers within the participants, there could be a small number bias. 

Overall, female and non-binary people, postdocs, and people from under-represented communities feel more left out compared to their colleagues. People who belong to several of the categories are even more impacted, for instance among female postdocs, more than one third felt unfairly absent from co-author lists.

\textbf{$\rightarrow$ Recommendation 3: Being mindful and careful about whom to include in publications and projects and particularly discussing the implications with postdocs, female and non-binary people, and people from under-represented groups. For instance, this can imply a code of publishing conduct within consortia.}

\section{Disrespect}
\label{sec:Disrespect}

One main lesson learned from the previous survey was the need to explicitly define what is meant by the notion of ``disrespect" in the survey questions. In this new survey, we therefore specified: ``By disrespect, we mean light behaviors that can make you or somebody uncomfortable, for instance cutting off somebody while they are talking, talking over another person without their consent, downplaying someone’s idea, expecting social role from younger people or women (taking notes, planning social events...). More reprehensible behaviors will be the purpose of the next section." 

Questions 17 to 21 of the survey (see Appendix \ref{sec:AppendixA}) addressed participants' feeling of having been disrespected under this definition. We evaluate this notion with two statuses: 1) victim: have people experienced disrespect and as victims, did they receive support from peers?, 2) witness: have people witnessed disrespect and as witnesses, how did they behave? Note that Sec. \ref{sec:Allies} is specifically dedicated to a deeper analysis of allyship, which includes the notion of witness. 

It should also be mentioned that even if this section of the survey questions was titled ``disrespect", and question 17 used the term ``disrespect'', question 19 was phrased as ``inappropriate behavior" instead of ``disrespect", which might have impacted some responses.

\begin{figure*}
   \begin{center}
   \begin{tabular}{ccc}
   \includegraphics[width=5.2cm]{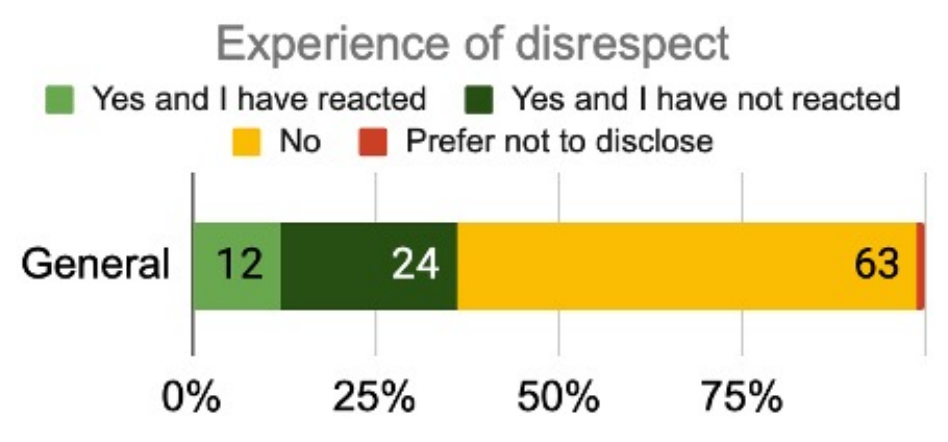} & \includegraphics[width=5.2cm]{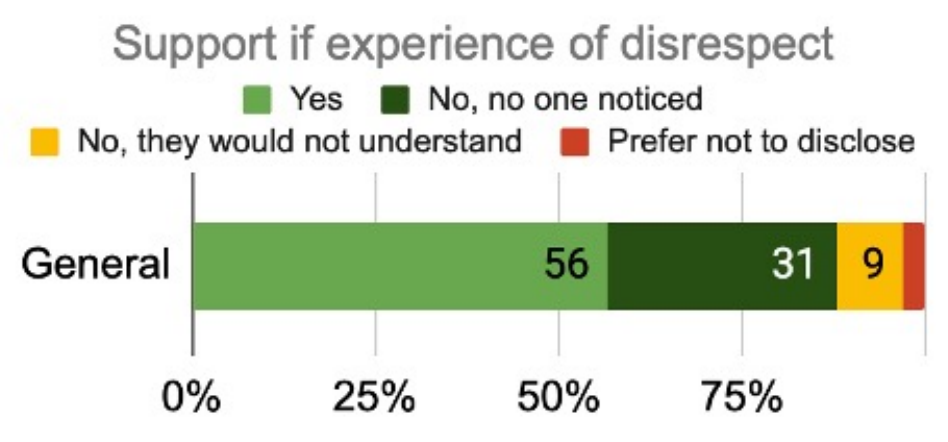} & \includegraphics[width=5.2cm]{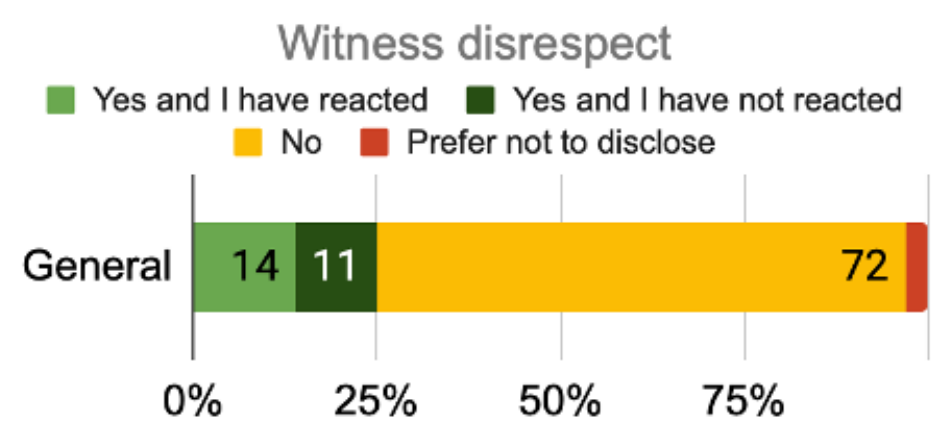} \\
   \includegraphics[width=5.2cm]{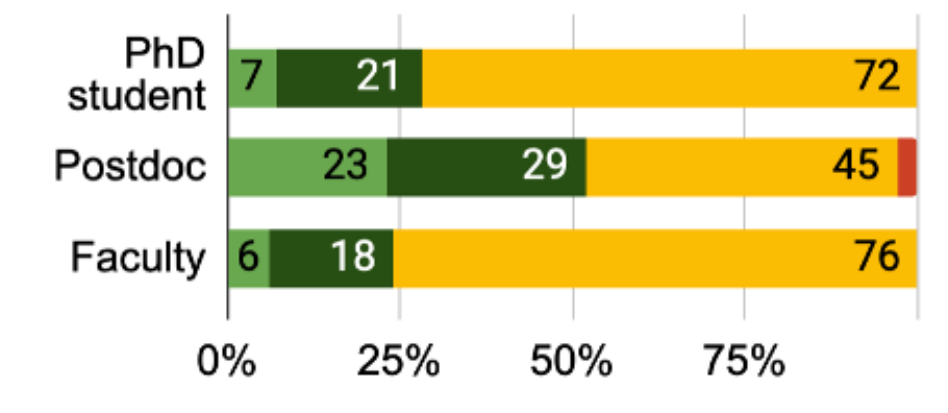} & \includegraphics[width=5.2cm]{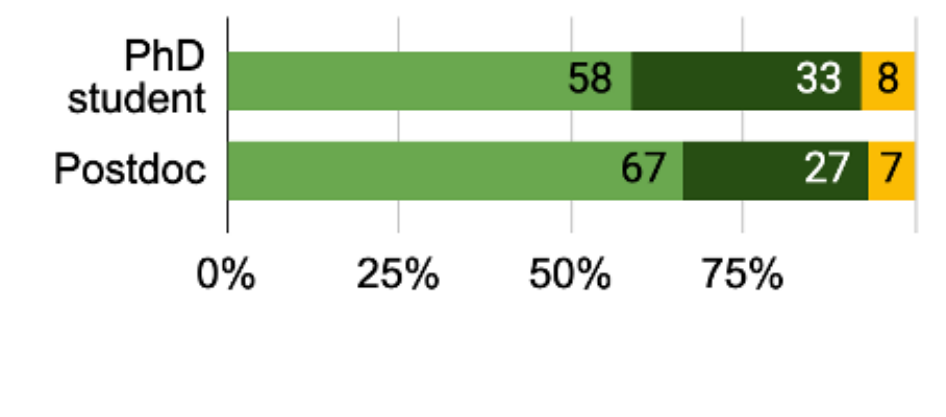} & \includegraphics[width=5.2cm]{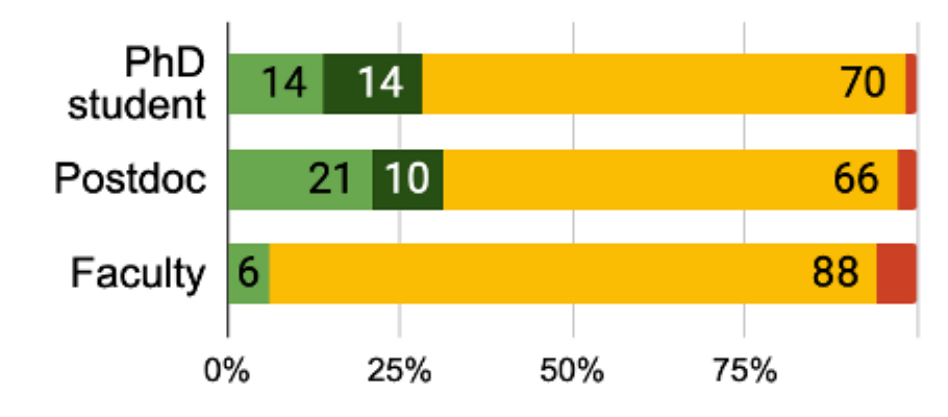} \\
   \includegraphics[width=5.2cm]{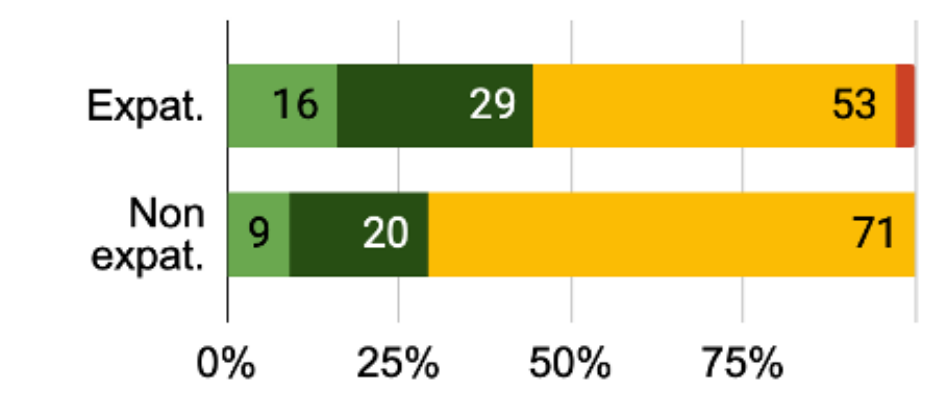} & \includegraphics[width=5.2cm]{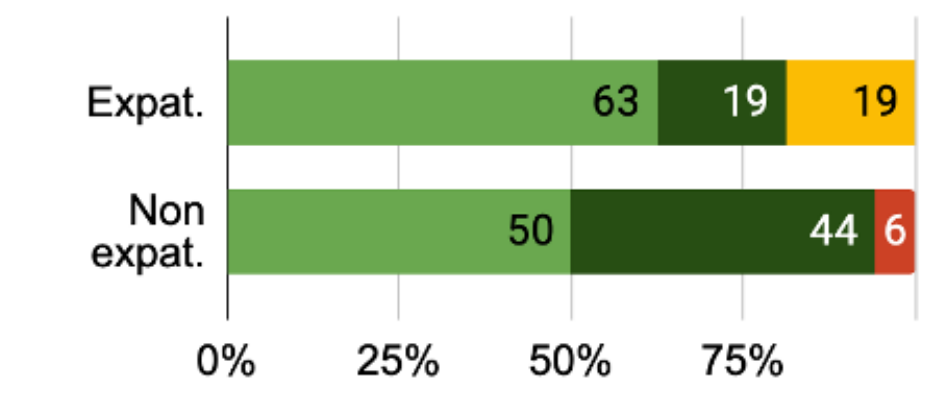} & \includegraphics[width=5.2cm]{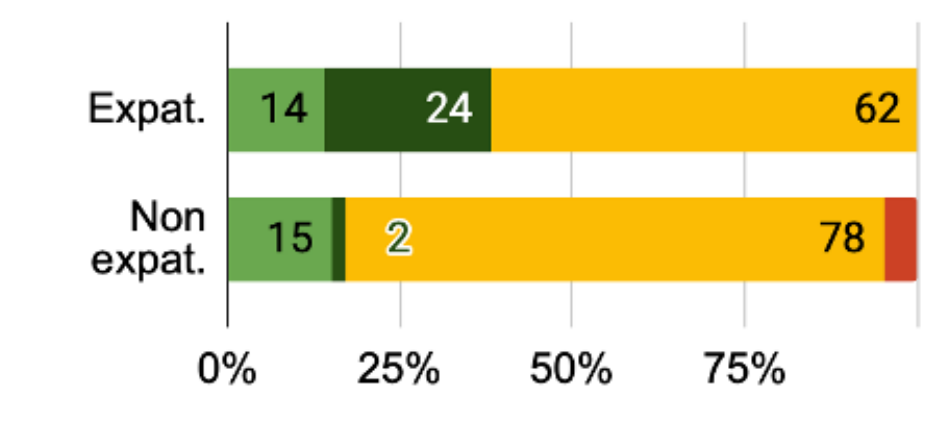} \\
   \includegraphics[width=5.2cm]{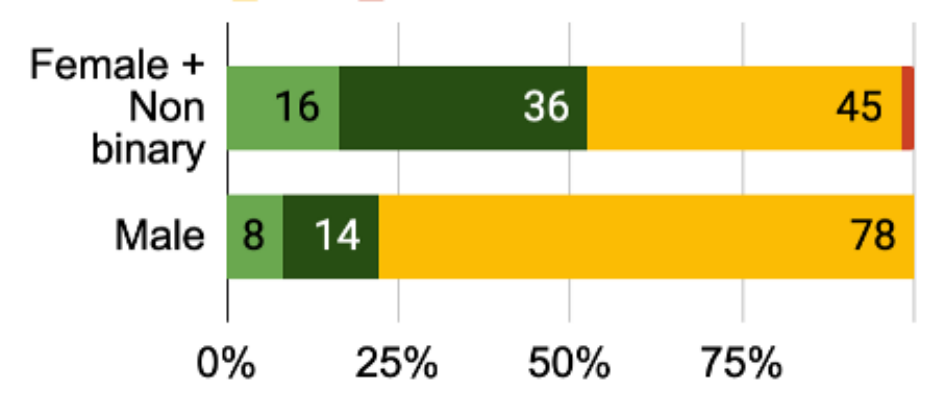} & \includegraphics[width=5.2cm]{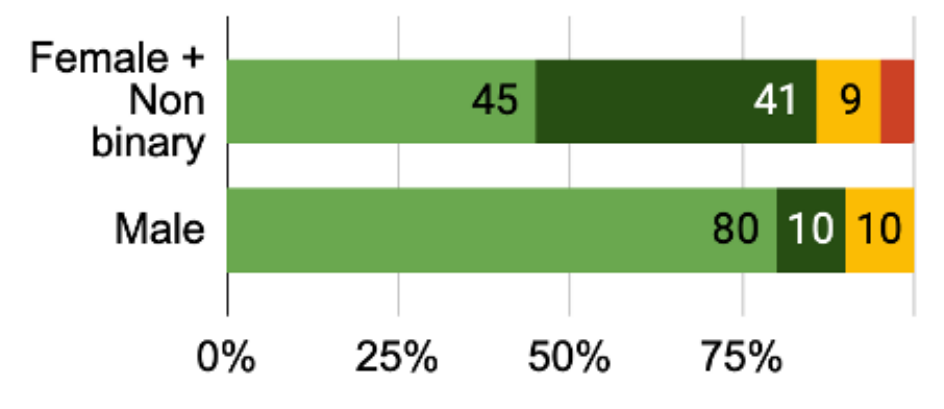} & \includegraphics[width=5.2cm]{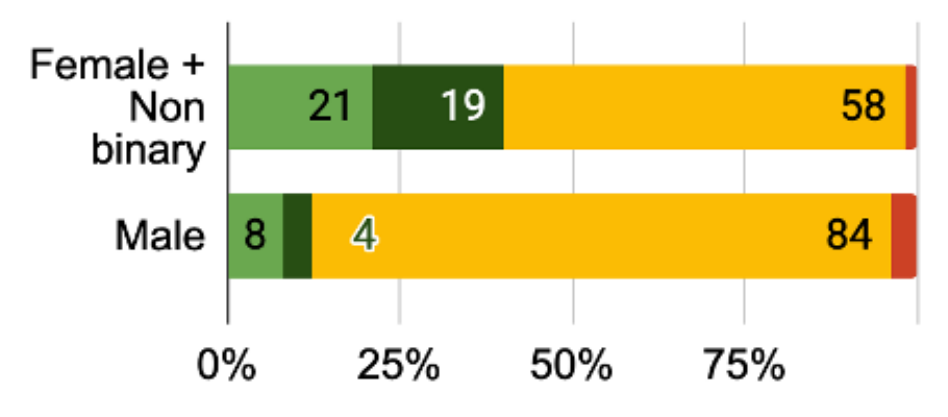} \\
   \includegraphics[width=5.2cm]{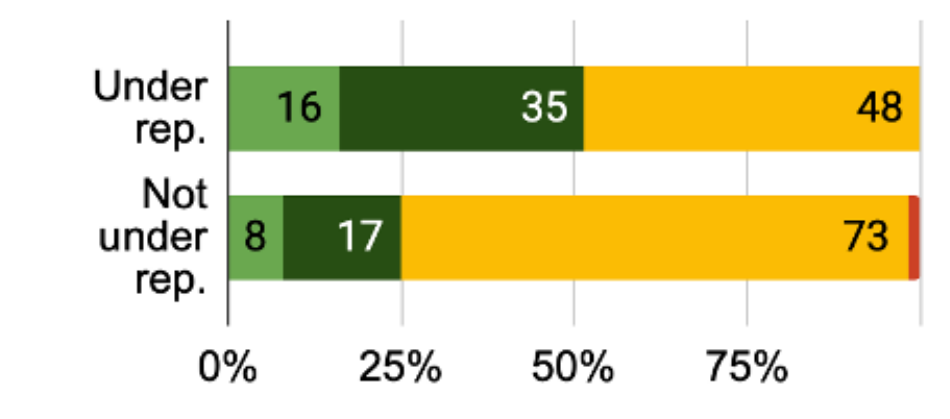} & \includegraphics[width=5.2cm]{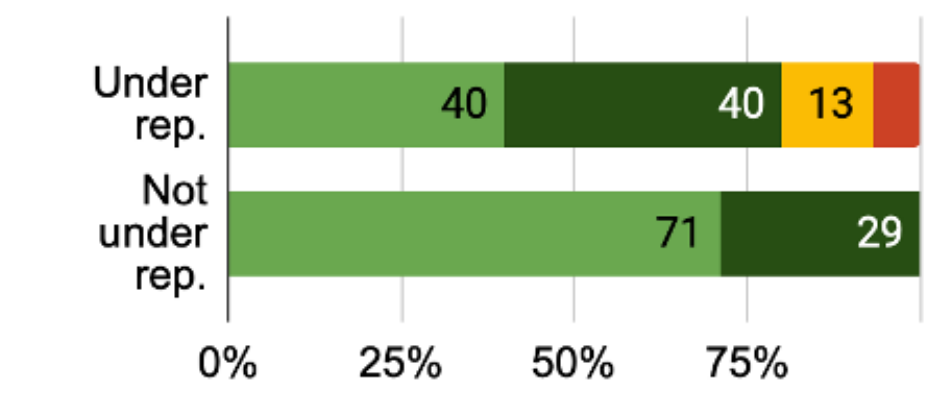} & \includegraphics[width=5.2cm]{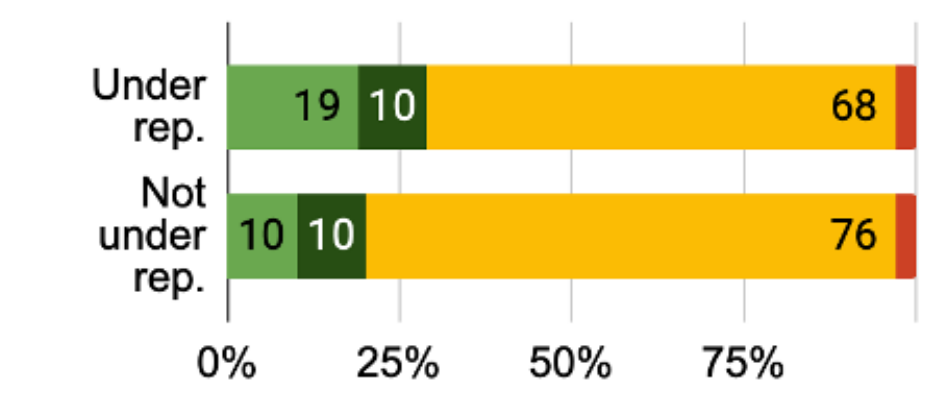} \\
   \includegraphics[width=5.2cm]{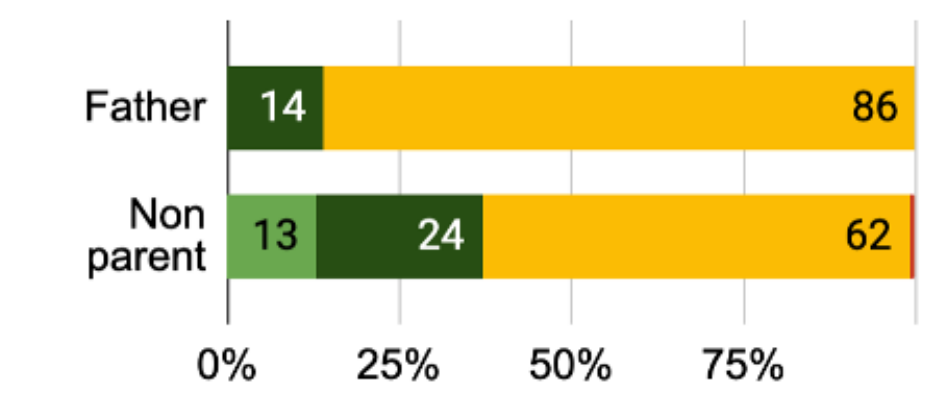} & \includegraphics[width=5.2cm]{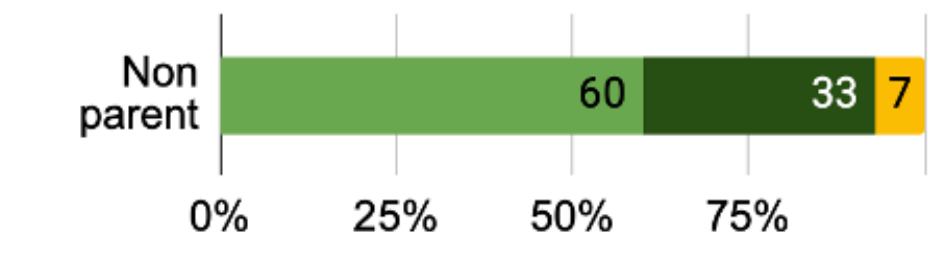} & \includegraphics[width=5.2cm]{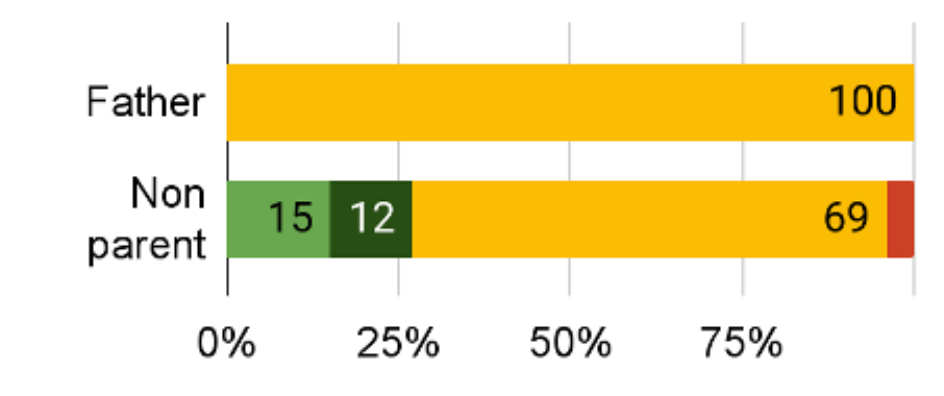} \\
   \end{tabular}
   \end{center}
   \caption 
   { \label{fig:Sec5_Fig1} 
Outcome of the questions regarding disrespect since January 2020, for both victims and witnesses: (left) percentages of people, per category, having experienced a situation of disrespect and their possible reaction, (center) percentages of victims of disrespect, per category, having been supported by other people, (right) percentages of people, per category, having witnessed a situation of disrespect and their possible reaction. Data including too small samples (less than 6 people) have been excluded from this graphics.} 
   \end{figure*}

Figure \ref{fig:Sec5_Fig1} provides the main quantified outcomes of the questions regarding disrespect, both for victims and witnesses. It presents the percentages of respondents, in general and per category, who since January 2020: (left) experienced a situation of disrespect, (center) received support from other people when they experienced disrespect, and (right) witnessed a situation of inappropriate behavior. For the first and third questions in this section, the response options allowed participants to indicate whether they reacted to the disrespect, which is addressed in more detail by later questions. The data for some categories (faculty and fathers) are missing in the second column of this figure due to small numbers (1 father and 3 faculty people concerned). 


$\bullet$ \textbf{Results per career stage:} The outcomes for postdocs are concerning: $52\%$ of them declare having experienced a situation of disrespect in the last $2.5$ years. They are also more likely to react when confronted with disrespect, both as a victim or a witness. Faculty are the least likely to witness or be aware of disrespect.

$\bullet$ \textbf{Results per expatriate status:} Expatriate participants are more often victims of disrespect, with a rate of $45\%$, and notice more when witnessing a situation of disrespect ($38\%$). This could be related to the correlation between expatriation and job position, since expatriates tend to be postdoctorate researchers.

$\bullet$ \textbf{Results per gender:} Along with postdocs, women and non-binary people disproportionately experience disrespect ($52\%$), and receive less support from other people when they are disrespected. As witnesses, they are the most likely to notice ($40\%$) and react ($21\%$) to disrespect.

$\bullet$ \textbf{Results according to community representation:} Participants from under-represented groups experience by far ($51\%$) more disrespect than their counterparts, receive the least support when they are a victim of disrespect ($40\%$), and are also more likely to react when they witness disrespect. 

$\bullet$ \textbf{Results for fathers vs non parents:} It seems that fathers are less likely to experience and notice situations of disrespect. However, the effect of fatherhood on situations of disrespect, both as victims and witnesses, is more complex to analyze due to the small numbers and the high correlation with the job position.

Overall, $36\%$ of participants declared having experienced disrespect since January 2020, and more than half of the victims received support from other people. We also note that disrespect is more experienced by postdoctorate researchers, female and non-binary people, and participants from under-represented groups. In general, one quarter of people notice when a situation of disrespect happens (mostly expatriates, females and non-binary people). However, expatriates are less likely to react, which can be correlated with occupying a non-permanent job which equates to a more vulnerable career position.

We should be particularly careful for the well-being of people who accumulate several discrimination factors: for instance $67\%$ of female postdocs and $55\%$ of expatriates from under-represented groups have experienced disrespect since January 2020. In addition, among those who did not experience disrespect for the last $2.5$ years, only $10\%$ have noticed, as witnesses, a situation of disrespect, while among the people who did experience disrespect, the percentage of witnesses went up to $50\%$: the people most likely to witness disrespect are people belonging to groups that are most disrespected, indicating a lack of awareness in the other groups.

In this section of the survey, there are also two open questions that aimed at learning 1) how witnesses of disrespect reacted (if they did), and 2) why they did not react (if they did not). Several reactions or strategies to react to disrespect were reported: 

$\bullet$ during the situation of disrespect: 1) reacting verbally, openly pointing out the problem, 2) redirecting the conversation, 3) being careful that the abuser does not receive more attention or recognition than the victim, and 4) being careful that the victim is referred to or recognized in a project.

$\bullet$ after the situation of disrespect: 1) providing emotional support to the victim, 2) asking for help from senior people or men, 3) more generally, consulting with peers or any relevant committee to evaluate the situation and what a proper reaction would have been, and 4) reporting the issue.

In case of no reaction, several reasons were put forward: 

$\bullet$ real-time uncertainties: 1) realizing only in retrospect, 2) not being sure of the intention/interpretation of the comments, or 3) no clear target.

$\bullet$ interpersonal dynamics: 1) being in a hierarchic or power dynamic with the abuser (permanent/non permanent), 2) too many people around, or 3) fear of judgement/reputation, of being annoying.

$\bullet$ another witness reacted (including the supervisor).

Witnesses have a crucial role in such situations, however power dynamics make the situation very uncomfortable or dangerous to react to for certain witnesses or even victims (early-career researchers for instance). 

\textbf{$\rightarrow$ Recommendation 4: Spreading or acquiring knowledge, for instance through training or talks, to be able to interpret a possible situation of disrespect and evaluate the well-being of people. This could target everybody, and particularly people to refer to in case of issues, such as supervisors, men, senior researchers, and others who may be less vulnerable and in a better position to help (specifically from the perspective of people from traditionally discriminated groups). This could help everyone understand blind spots and acquire the ability to isolate the real situation with subjective biases one may have.}

\textbf{$\rightarrow$ Recommendation 5: Spreading or acquiring knowledge, for instance with training or talks, to be able to react properly when confronted with a possible situation of disrespect: appropriate behavior towards the possible victim, people to contact (advisor, anti-harassment cell, human resources, institute's equity committee...), revisit existing procedures, etc. This could target everybody, and particularly people to refer to in case of issues, such as supervisors, men, senior researchers, and others who may be less vulnerable and in a better position to help.}

\textbf{$\rightarrow$ Recommendation 6: When witnessing a situation of disrespect, first always checking with the possible victim about how they feel and their needs. It could also help to verify our own interpretation of the situation.}

\section{Inappropriate behaviors}
\label{sec:Inappropriate behaviors}

As in the previous section, we added a definition of inappropriate behavior to the 2022 survey which was not done in the 2019 one: ``By inappropriate behaviors, we mean any social behavior that can make you uncomfortable even if it is not necessarily legally reprehensible. Here are a few specific examples of inappropriate behaviors: condescending remarks, discriminating behavior based on ethnicity, inappropriate jokes, racist jokes, staring at, sexual remarks or questions at a work environment, disrespect based on one's culture and identity, ignoring or excluding somebody during a meeting, preventing somebody from attending meetings, remarks on parental leaves, sexual harassment, bullying...".

\begin{figure*}
   \begin{center}
   \begin{tabular}{ccc}
   \includegraphics[width=5.2cm]{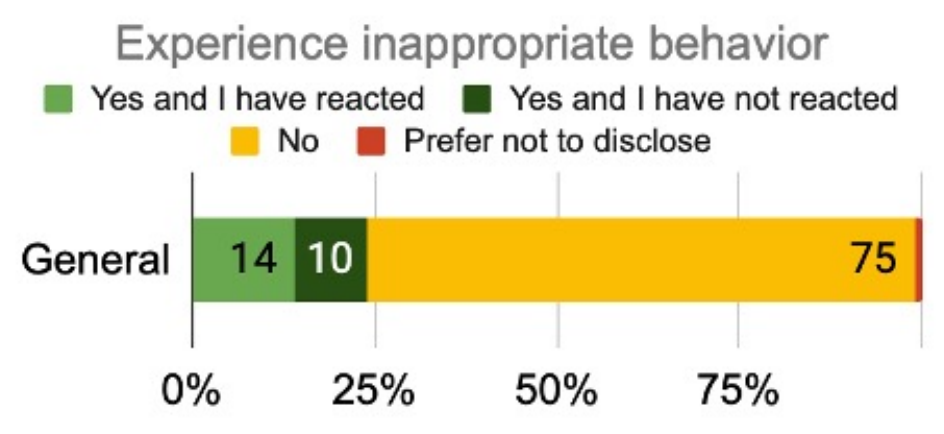} & \includegraphics[width=5.2cm]{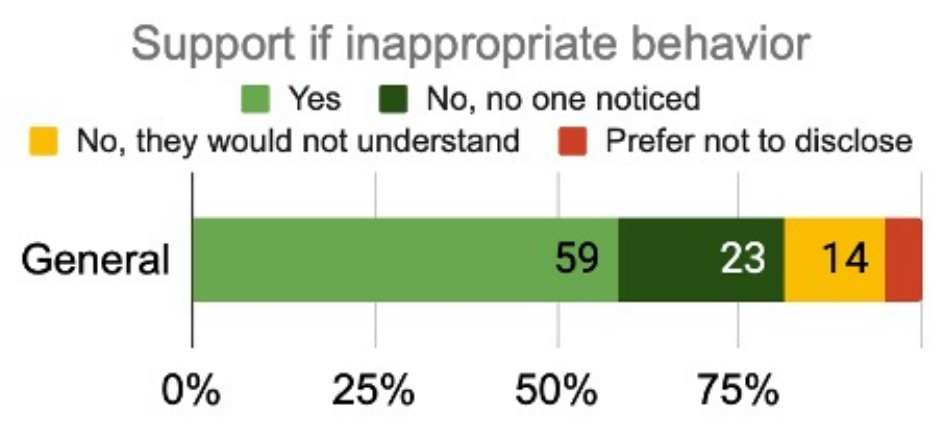} & \includegraphics[width=5.2cm]{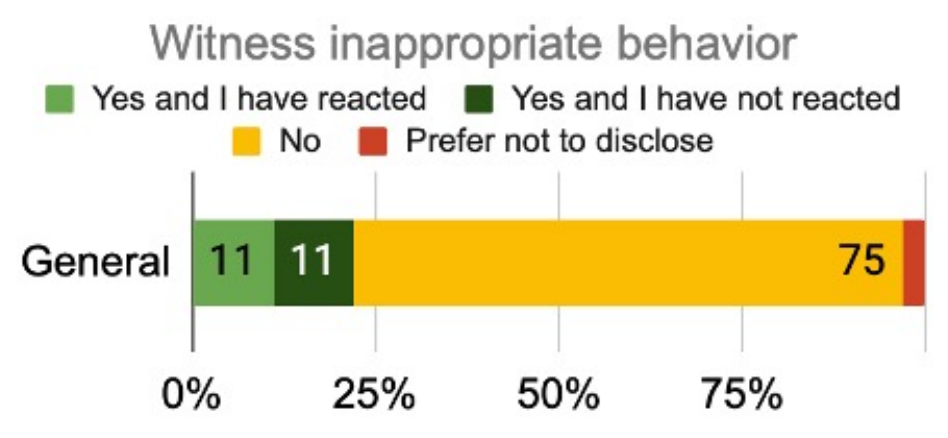} \\
   \includegraphics[width=5.2cm]{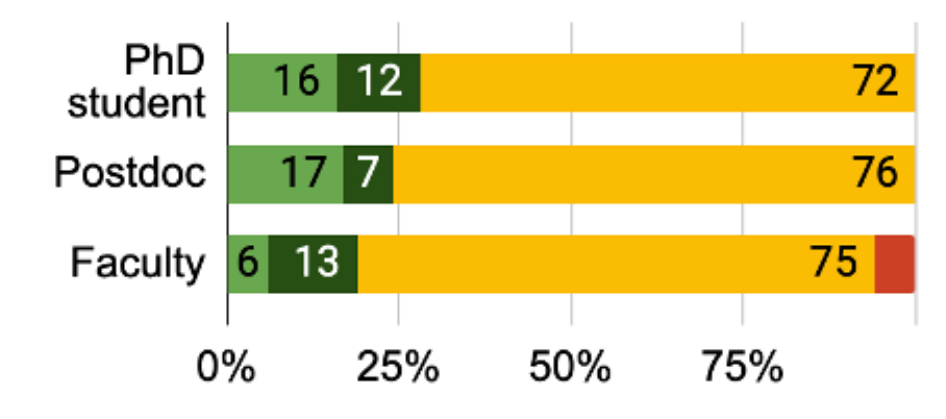} & \includegraphics[width=5.2cm]{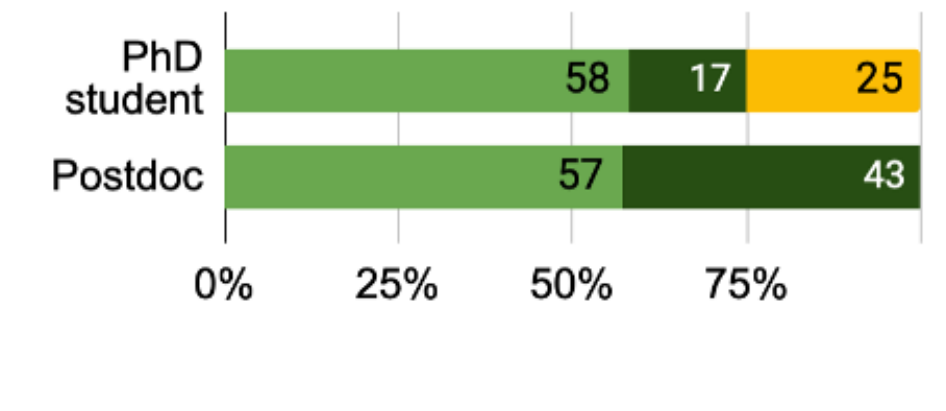} & \includegraphics[width=5.2cm]{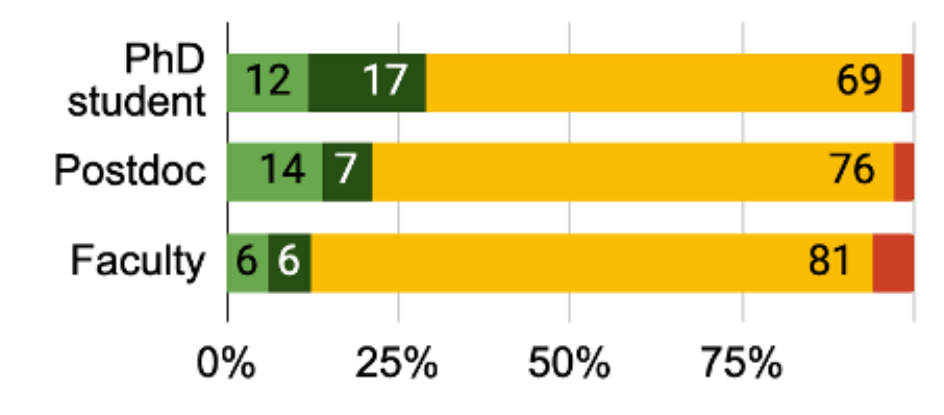} \\
   \includegraphics[width=5.2cm]{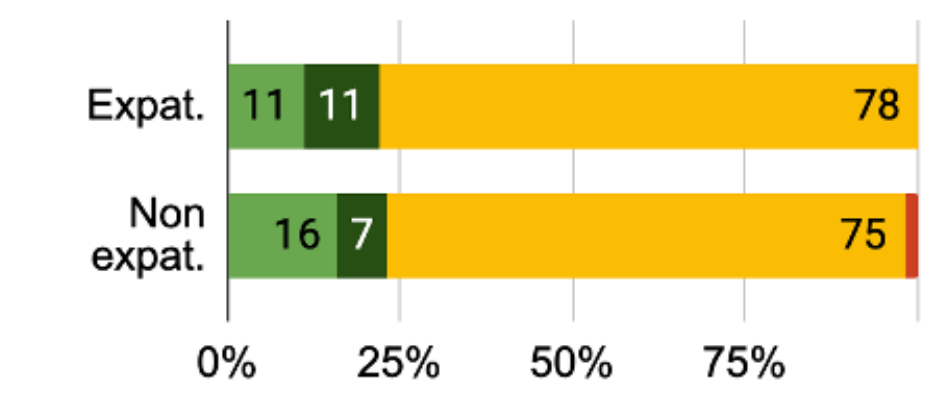} & \includegraphics[width=5.2cm]{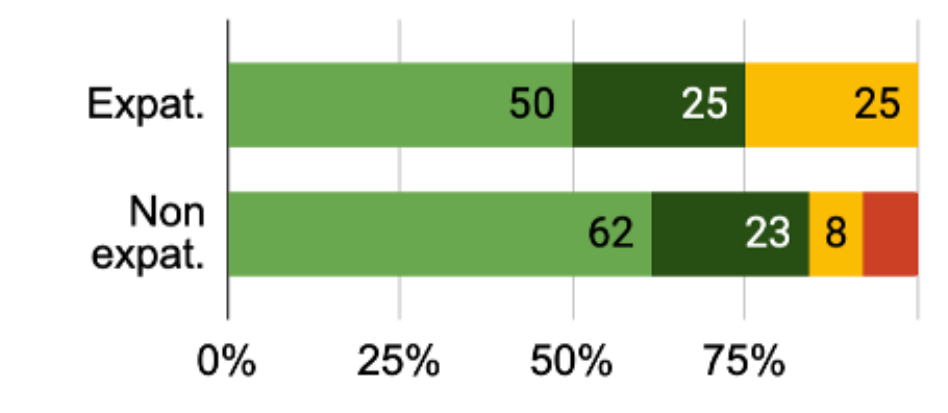} & \includegraphics[width=5.2cm]{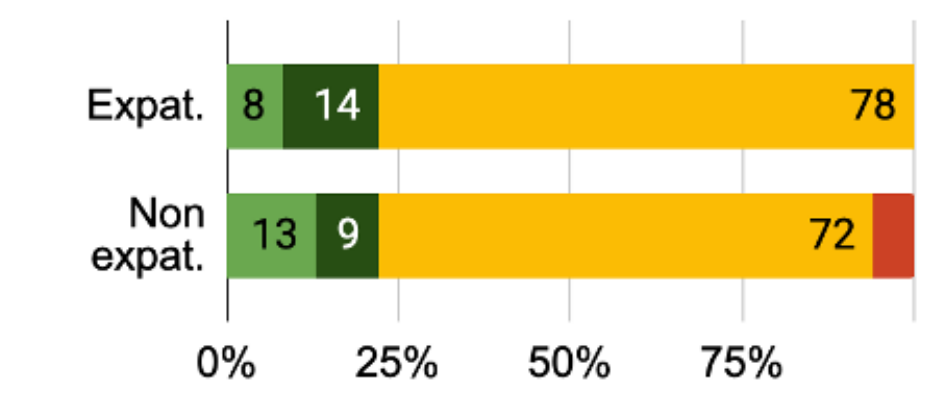} \\
   \includegraphics[width=5.2cm]{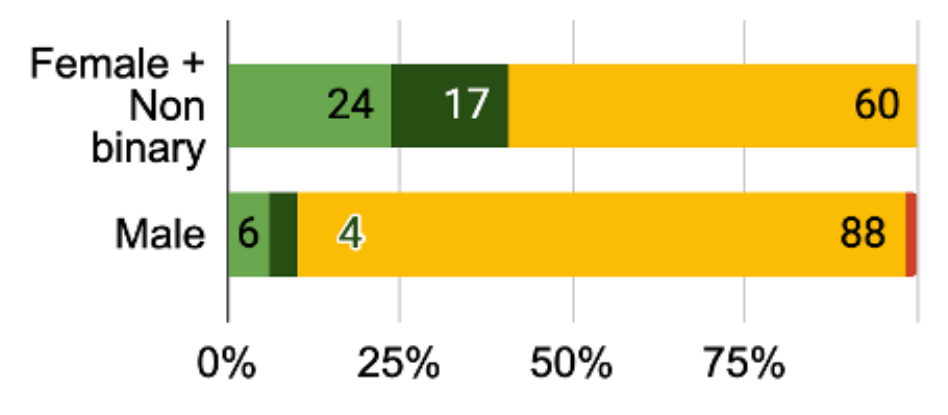} & \includegraphics[width=5.2cm]{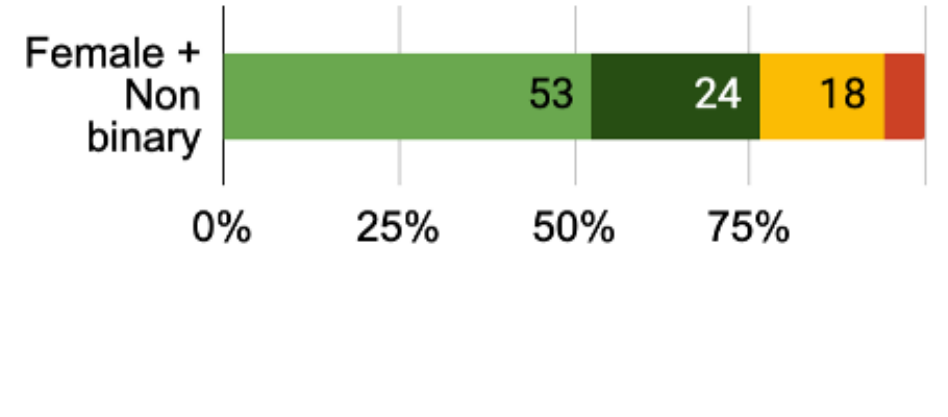} & \includegraphics[width=5.2cm]{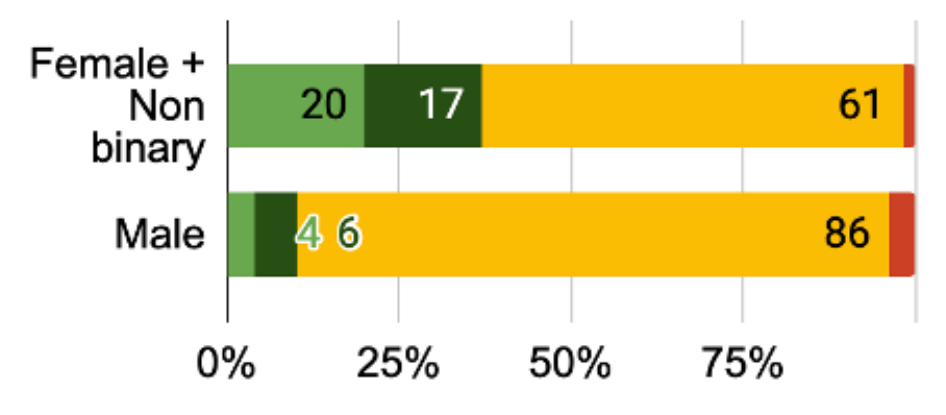} \\
   \includegraphics[width=5.2cm]{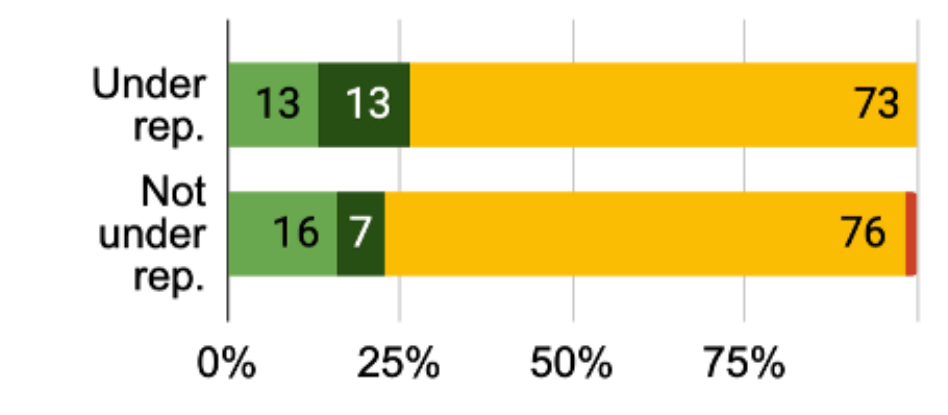} & \includegraphics[width=5.2cm]{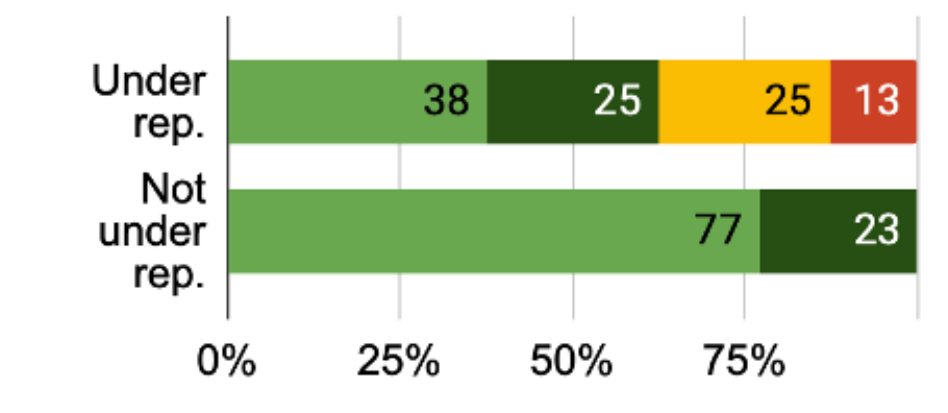} & \includegraphics[width=5.2cm]{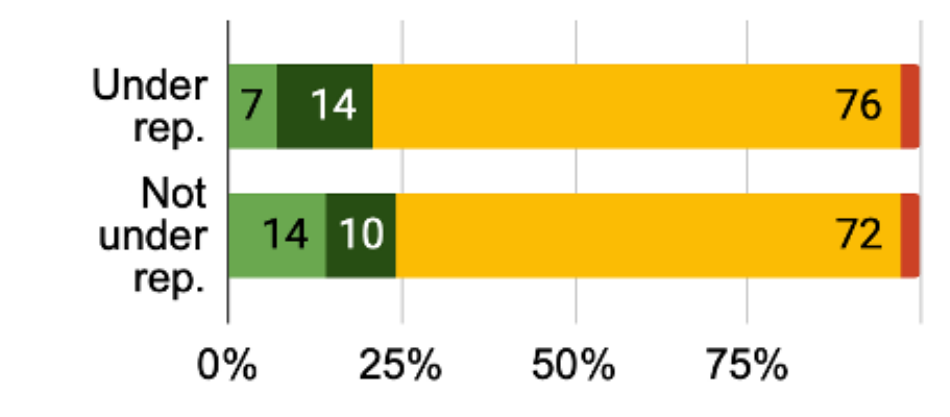} \\
   \includegraphics[width=5.2cm]{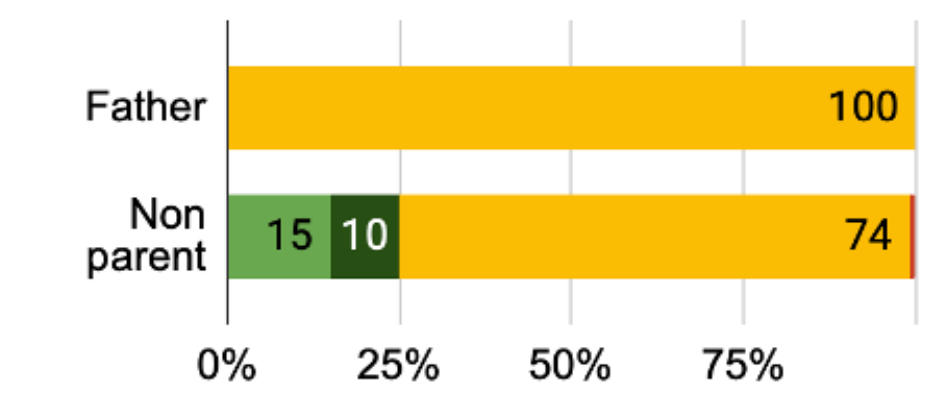} & \includegraphics[width=5.2cm]{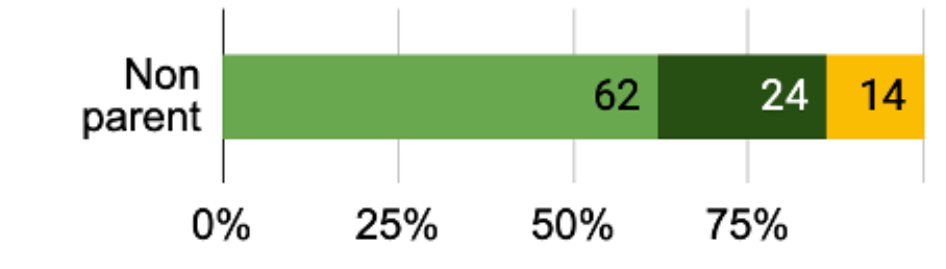} & \includegraphics[width=5.2cm]{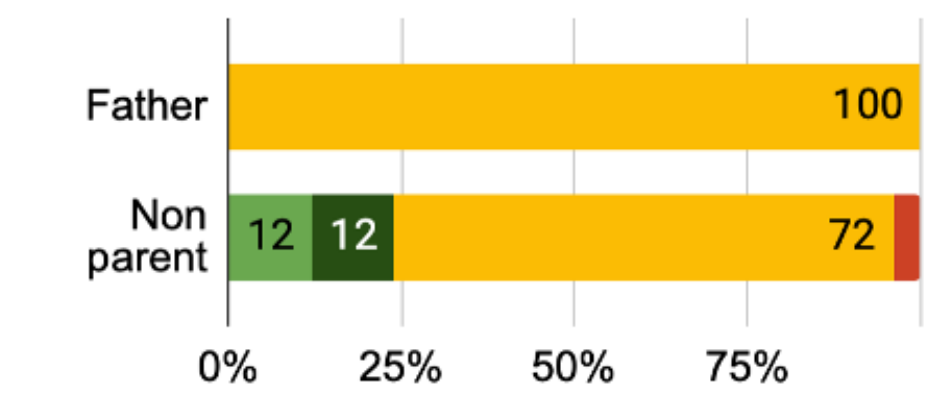} \\
   \end{tabular}
   \end{center}
   \caption 
   { \label{fig:Sec6_Fig1} 
Outcome of the questions regarding inappropriate behavior, with both the status of victim and the status of witness, all since January 2020: (left) percentages of people, per category, having experienced a situation of inappropriate behavior and their possible reaction, (center) percentages of victims of inappropriate behavior, per category, having been supported by other people, (right) percentages of people, per category, having witnessed a situation of inappropriate behavior and their possible reaction. Data including too small samples (less than 6 people) have been excluded from this graphics.} 
   \end{figure*}

This section corresponds to questions 22 to 26 of the survey (see Appendix \ref{sec:AppendixA}) which concern the experience of inappropriate behaviors for victims and witnesses, since January 2020. Figure \ref{fig:Sec6_Fig1} presents the main outcomes with (left) the percentages of people, per category,  having experienced a situation of inappropriate behavior and their possible reaction, (center) percentages of victims of inappropriate behavior, per category, having been supported by other people, (right) percentages of people, per category, having witnessed a situation of inappropriate behavior and their possible reaction. Plots for small-number groups (5 people or less) have been removed.

In general, almost one quarter of participants have experienced inappropriate behaviors within a $2.5$ year range (versus $33\%$ in the 2019 survey but with no time restriction), and $59\%$ of victims got support from other people. In the role of witness, $22\%$ noticed situations of inappropriate behaviors ($48\%$ in the 2019 survey despite no time restriction: the 2019 question was about witnessing inappropriate behaviors ever). The phenomenon seems slightly less frequent than disrespect, but victims get equivalent support. However, these rates remain high and reveal an unsafe environment.

$\bullet$ \textbf{Results per career stage:} Non-permanent researchers and especially PhD students are more often victims ($28\%$ for PhD students) and witnesses ($29\%$ for PhD students) of inappropriate behaviors. In 2019, postdocs had declared having experienced inappropriate behaviors twice ($56\%$) more than PhD students ($21\%$) and faculty researchers ($28\%$). The added time restriction for the new survey might have removed or reduced the bias towards PhD students who have a shorter career. However, the fact that the earlier in their career the researcher is, the more exposed they are to inappropriate behavior is highly concerning.

$\bullet$ \textbf{Results per expatriate status:} No clear trend appears within these categories. 

$\bullet$ \textbf{Results per gender:} Female and non-binary participants are generally both more often victims ($41\%$, versus $52\%$ in 2019 but for women only and without time restriction) and witnesses ($37\%$, versus $55\%$ in 2019 but for women only and without time restriction) of inappropriate behaviors than their male colleagues. This is especially noteworthy given that according to \cite{DOrgeville2014}, inappropriate behavior is one of the main reasons why women leave academia.

$\bullet$ \textbf{Results according to community representation:} There is no clear trend for experiencing or witnessing situations of inappropriate behaviors in these categories, but participants that do not belong to under-represented groups get twice more support than their colleagues from under-represented groups.

$\bullet$ \textbf{Results for fathers vs non parents:} Due to the small number of parents and the correlation with job position, this categorization is more complex to evaluate, but no father seems to have experienced or witnessed inappropriate behaviors since 2020.

This data quantifies an ongoing issue in our field, that specifically targets non-permanent researchers and female and non-binary people even if all categories are vulnerable. In addition, more people indicated that, as victims, they did not get support since witnesses did not notice or would not understand the ongoing inappropriate behavior, which indicates a general lack of information for witnesses and supervisors. 

As in the previous section on disrespect, people with intersectionality are even more impacted by inappropriate behaviors: for instance, female and non-binary PhD students experience ($36\%$) and notice (also $36\%$) situations of inappropriate behavior. In addition, the correlation between having experienced and witnessed such a situation is even higher than in the previous section: among people who did not experience inappropriate behaviors, $4\%$ have indicated having witnessed one (none of them reacted), while among people who experienced inappropriate behaviors, $77\%$ noticed such a situation as witnesses.


As witnesses, participants could share in the survey their reactions to inappropriate behaviors (if they reacted) or the reasons why they did not react. We summarize them below. 

Strategies when confronted with inappropriate behavior include:

$\bullet$ during the situation: calling attention through instantaneous public reaction, direct or humorous.

$\bullet$ after the situation: 1) checking with the victim, 2) asking for a senior researcher to react, 3) discussing among peers to evaluate actions to take, and 4) reporting.

Reasons that prevented witnessed from reacting include:

$\bullet$ external barriers: 1) online event, 2) language difficulties.

$\bullet$ someone else reacted (including the supervisor).

$\bullet$ fear of judgment and isolation.

$\bullet$ lack of information about what to do or whom to talk to.

In addition to extending  the recommendations described in Sec. \ref{sec:Disrespect} to incorporate inappropriate behaviors, we recommend:

\textbf{$\rightarrow$ Recommendation 7: Improving the safety of the workplace for everybody, in particular for early career scientists to preserve the well-being and overall inclusion and diversity in the field by helping them stay safely in academia. This includes highlighting the procedures in case of inappropriate behaviors (victim or witness), for instance with training or posters.}

\textbf{$\rightarrow$ Recommendation 8: A dedicated training session on existing procedures could be offered at the start of the PhD, as junior scientists (including PhD candidates) appear to be the most likely to encounter inappropriate behavior.}

\section{Allies}
\label{sec:Allies}

In the survey, the notion of ``ally" is defined as follows: ``An ally is somebody who 1) acknowledges the difficulties faced by under-represented or discriminated groups they do not belong to and 2) supports them when necessary or asked to". The capacity of an ally to take action when witnessing a certain situation can be inhibited by the so-called bystander effect: people are less likely to intervene in a situation that requires help in the presence of other people \citep{Latane1970, Hortensius2018}. 

This section refers to questions 27 to 29 in the survey (see Appendix \ref{sec:AppendixA}). The objective was to evaluate the self-identification of people as allies and their wish for information. A third question aimed at quantifying the access to (or knowledge of) formal procedures in case of harassment in institutions.

\begin{figure*}
   \begin{center}
   \begin{tabular}{cc}
   \includegraphics[width=6cm]{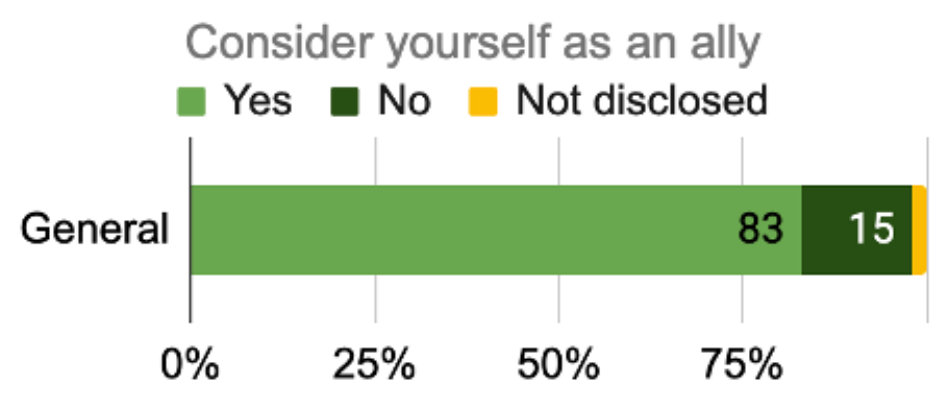} & \includegraphics[width=6cm]{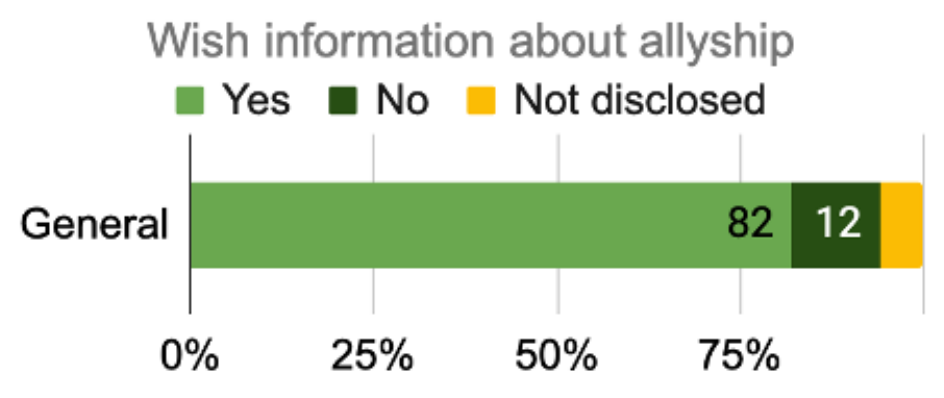} \\
   \includegraphics[width=6cm]{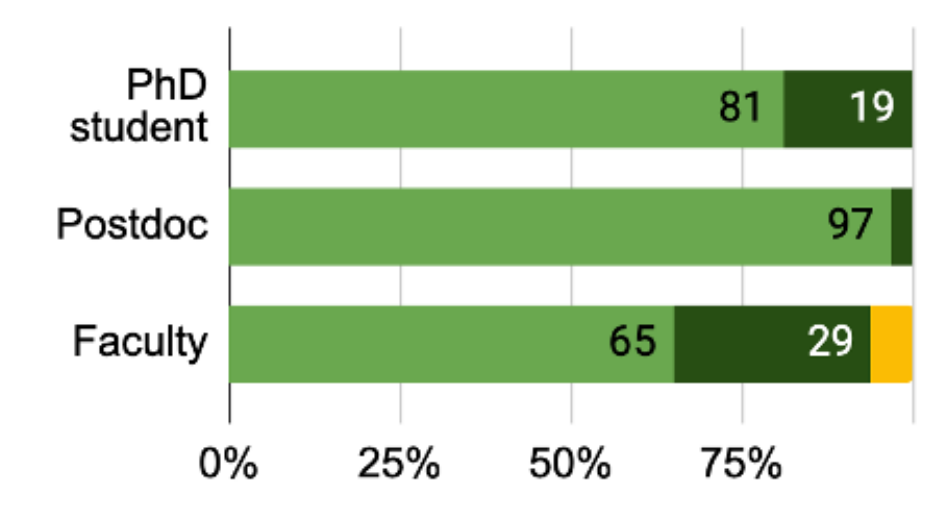} & \includegraphics[width=6cm]{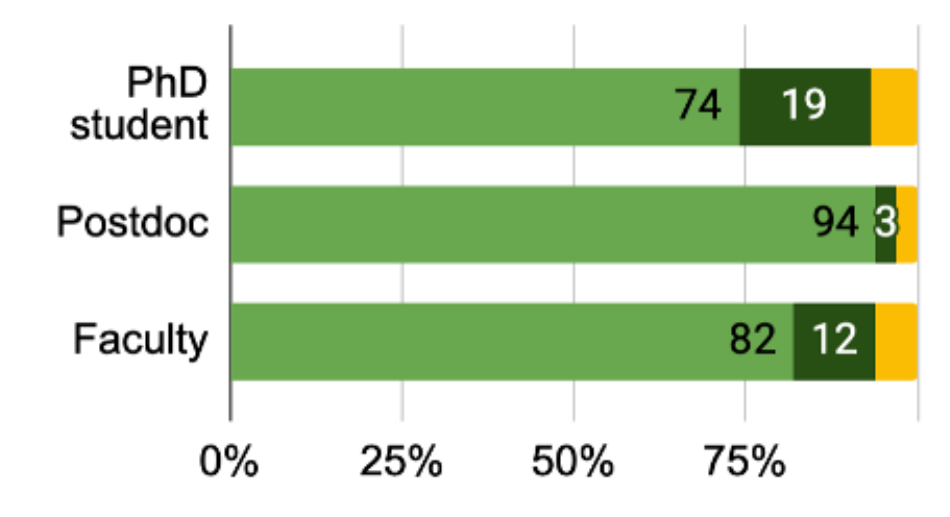} \\
   \includegraphics[width=6cm]{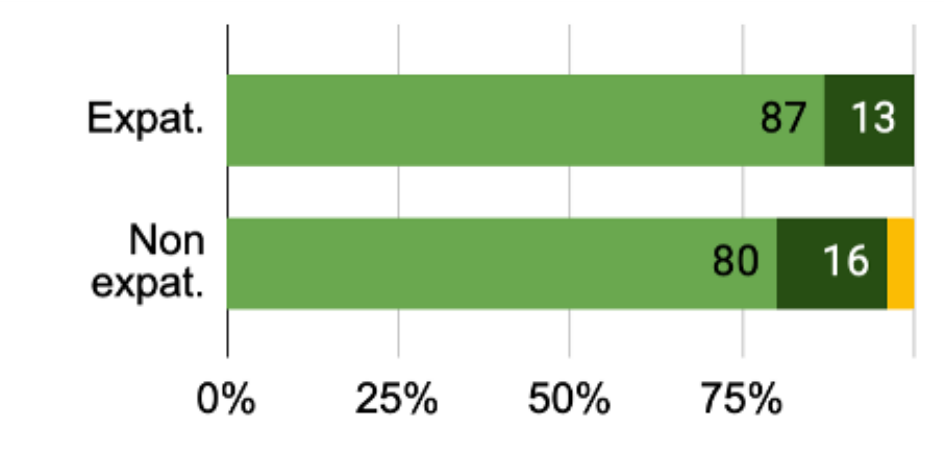} & \includegraphics[width=6cm]{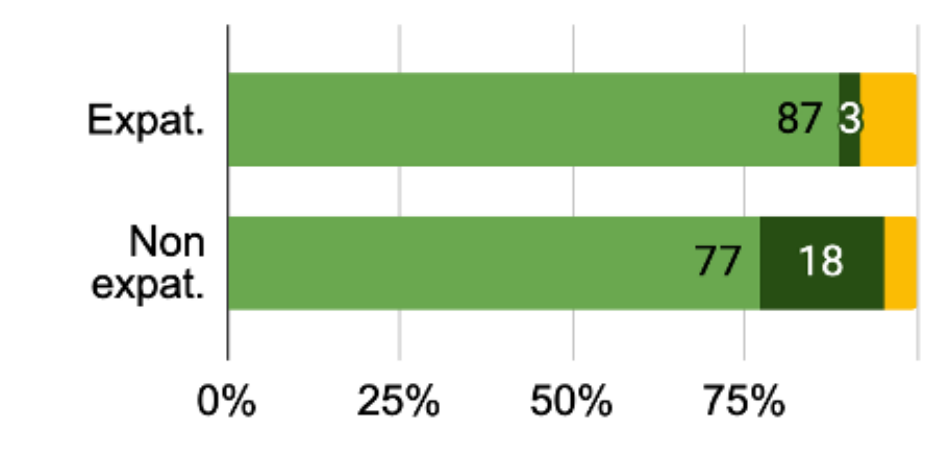} \\
   \includegraphics[width=6cm]{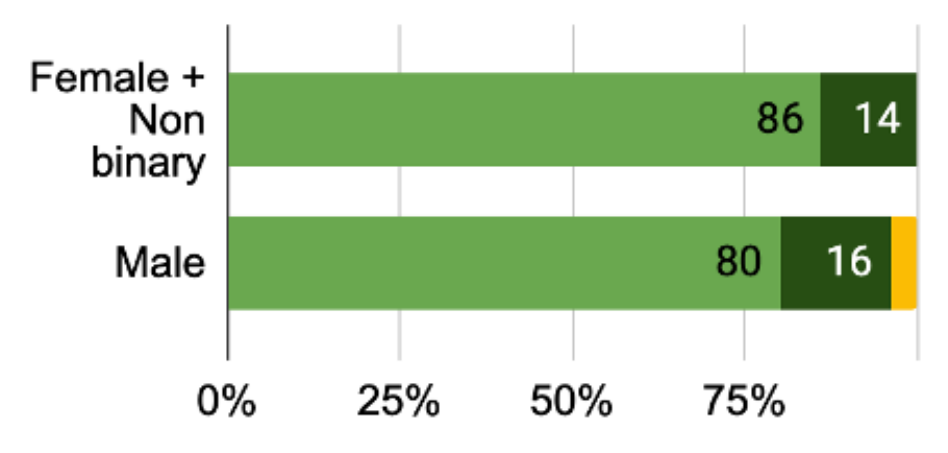} & \includegraphics[width=6cm]{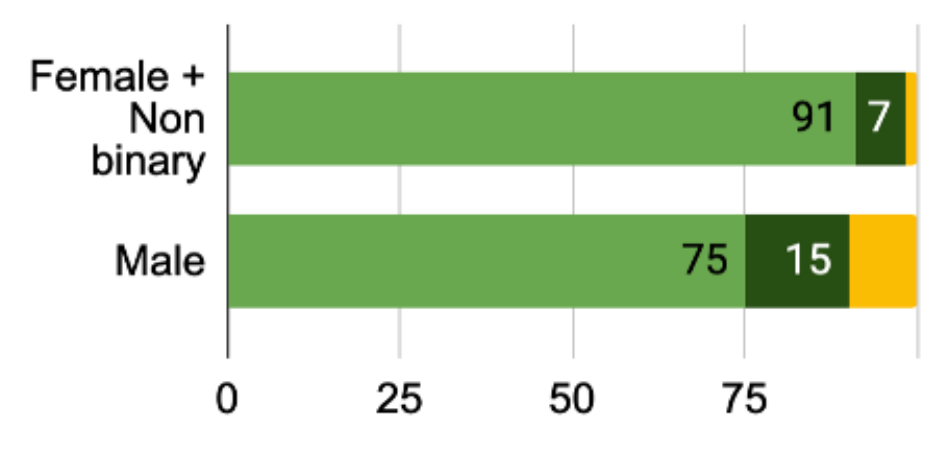} \\
   \includegraphics[width=6cm]{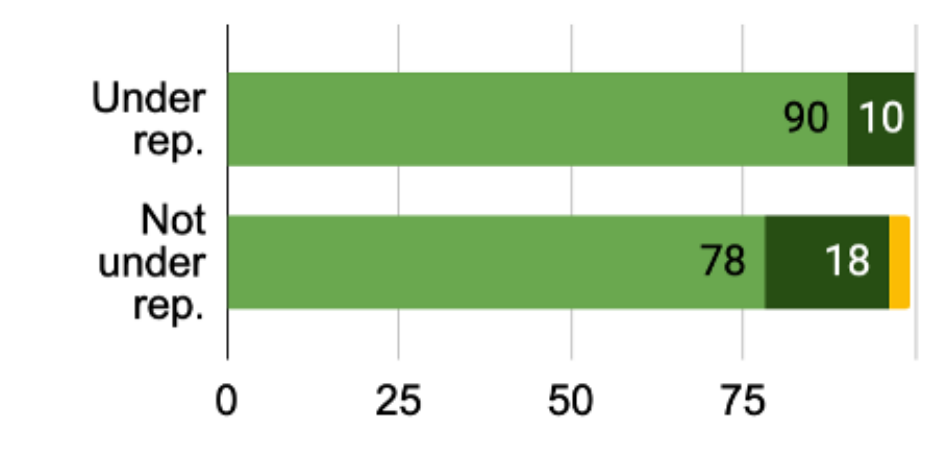} & \includegraphics[width=6cm]{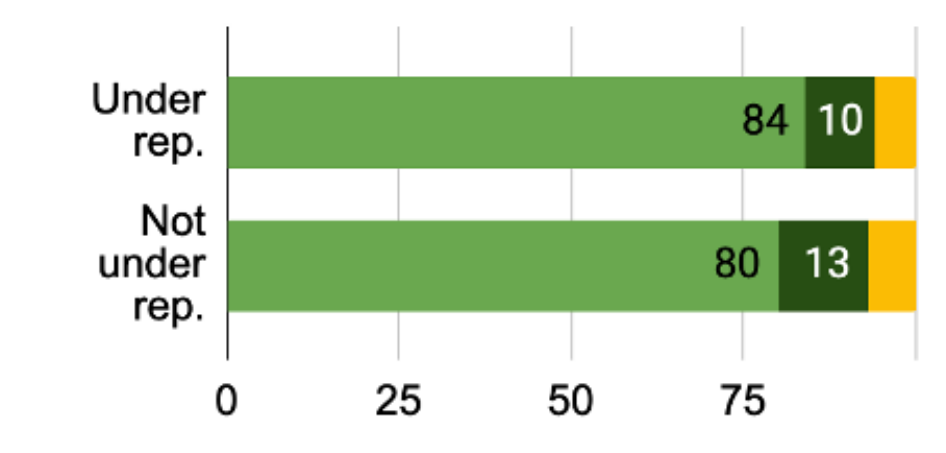} \\
   \includegraphics[width=6cm]{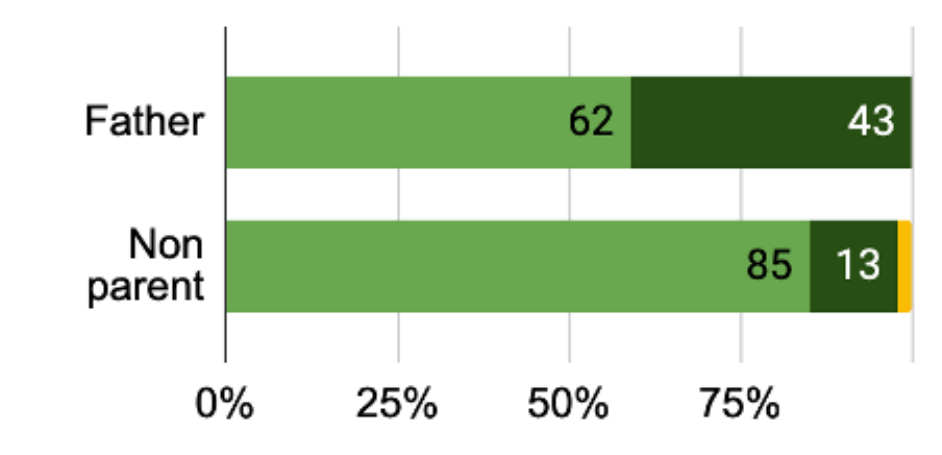} & \includegraphics[width=6cm]{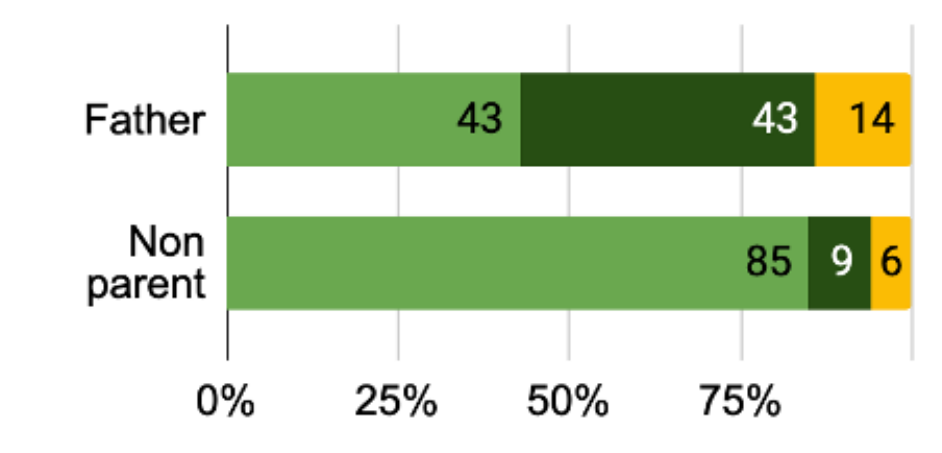} \\
   \end{tabular}
   \end{center}
   \caption 
   { \label{fig:Sec7_Fig1} 
Outcome regarding allyship: (left) percentages of respondents, in general and per category, who consider themselves as allies to under-represented groups, (right) percentages of respondents, in general and per category, who wish for talks or information about allyship.} 
   \end{figure*}

The results regarding self-identification as an ally and the wish for talks or information about allyship are presented in Figure \ref{fig:Sec7_Fig1}. Overall, $83\%$ of participants consider themselves as allies while $15\%$ do not. A large majority of people ($82\%$) ask for more information about allyship. 

$\bullet$ \textbf{Results per career stage:} Postdoctorate researchers are very likely to self-identify as allies ($97\%$) while faculty researchers are the least likely to identify as allies ($65\%$). However, the latter do seem to want to learn how to be allies ($82\%$), and postdocs are the group that most wish for information about allyship ($94\%$).

$\bullet$ \textbf{Results per expatriate status:} Expatriate respondents are more likely to want to learn how to be allies ($87\%$) than those who do not identify as expatriates ($77\%$).

$\bullet$ \textbf{Results per gender:} Female and non-binary respondents are more likely to want to learn how to be allies ($91\%$) than their male colleagues ($75\%$).

$\bullet$ \textbf{Results according to community representation:} People from under-represented groups are more likely to consider themselves as allies ($90\%$).

$\bullet$ \textbf{Results for fathers vs non parents:} Once again, we have very few respondents who are parents and a high correlation with a permanent position, which could explain the apparent trend for fathers to consider themselves as allies less often than non-parents. 

\begin{figure*}
   \begin{center}
   \begin{tabular}{cc}
   \includegraphics[width=7.3cm]{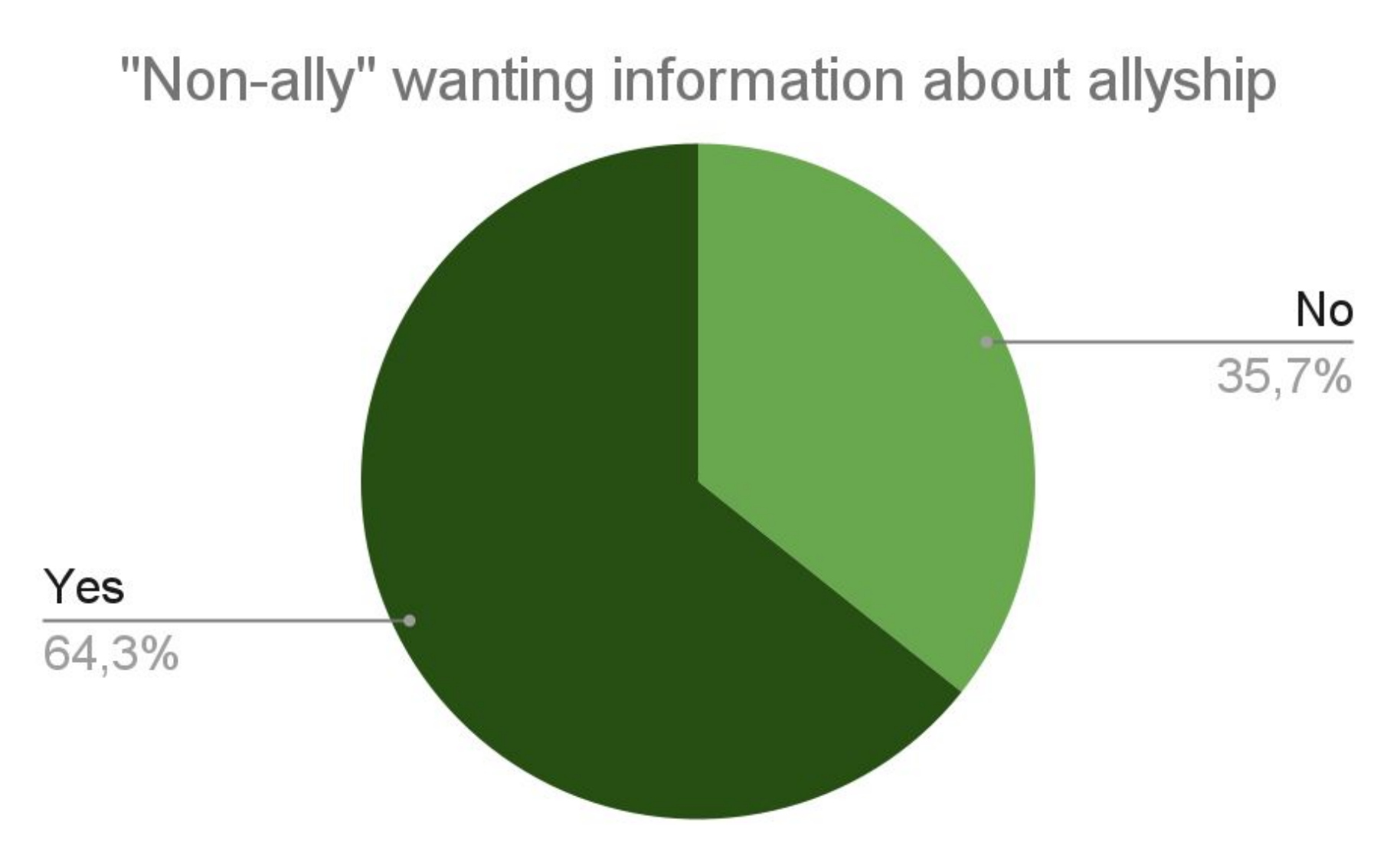} & \includegraphics[width=7.3cm]{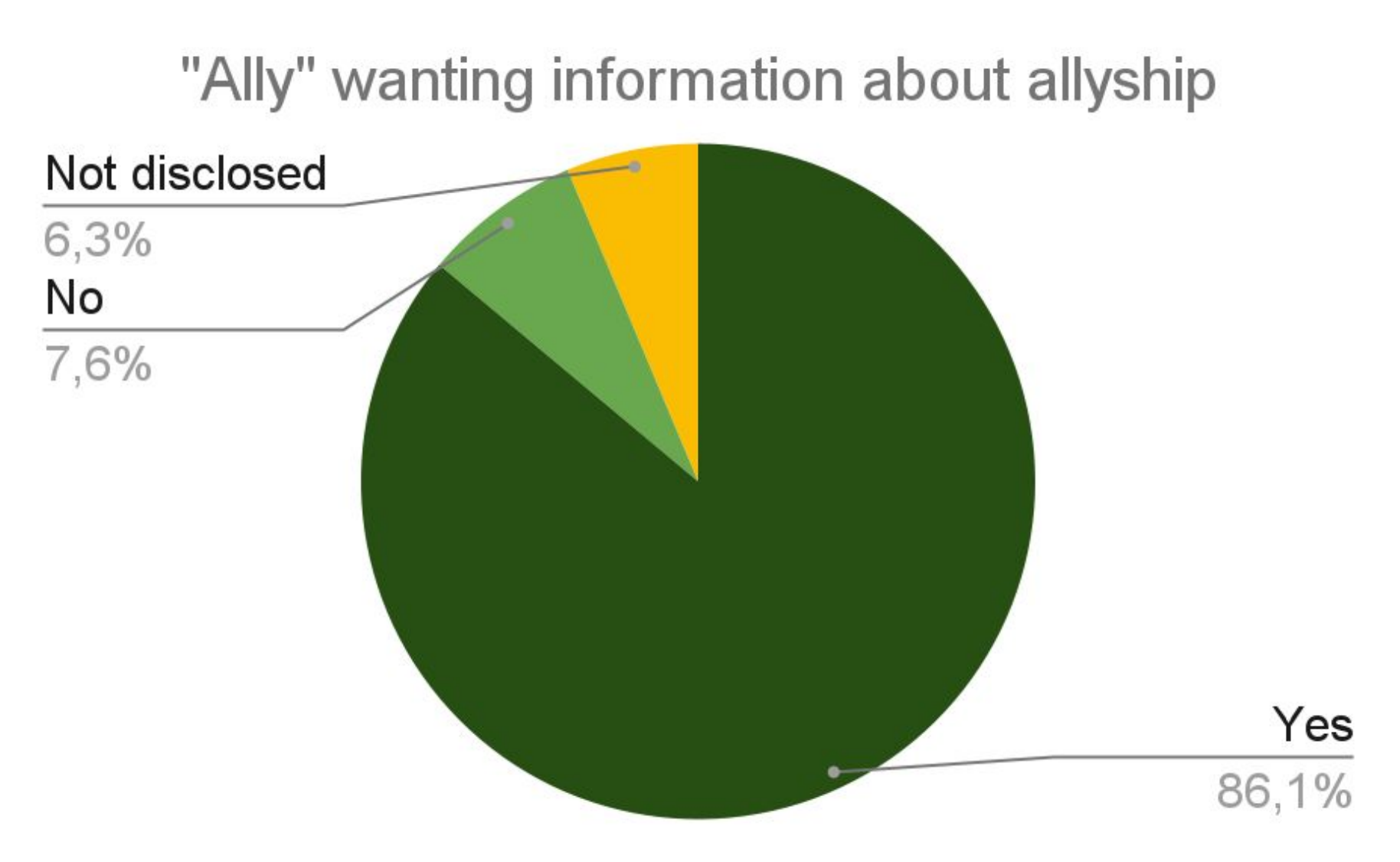} \\
   \end{tabular}
   \end{center}
   \caption
   { \label{fig:Sec7_Fig2} 
Correlation between wish for information about allyship and the self-consideration as an ally among participants. ``Non-ally" means a respondent who does not consider themself an ally, and ``ally" a respondent who considers themself an ally.} 
   \end{figure*}

Figure \ref{fig:Sec7_Fig2} shows the link between considering oneself an ally and willingness to learn about allyship: among ``non-allies" (i.e., respondents who do not consider themselves as allies), $64\%$ want to learn how to be allies. This percentage goes up to $86\%$ among ``allies" (i.e., respondents who consider themselves as allies). In general, this indicates that there is clearly desire and room for improvement in allyship, with an obvious and quite largely accepted route (through talks and information), even among people who do not necessarily identify themselves as allies.

Concerning procedures in case of harassment, almost one quarter ($22\%$) of respondents are not aware of procedures at their institution, which is concerning. This result is also quite homogeneous among categories (between $73\%$ and $82\%$ for all categories), with no clear trend.

\textbf{$\rightarrow$ Recommendation 9: There is a strong request to get more information about allyship, so we would recommend organizing a talk at conferences about biases, discrimination and allyship. This is in line with recommendations 3 and 4 formulated earlier.}

\section{Conclusions}
\label{sec:Conclusions}

These survey and paper come as a follow-up and improved version of the survey from the 2019 Spirit of Lyot conference and the article published from its outcomes \citep{Leboulleux2020}. The objective of these actions is to measure, monitor over time, and communicate about the well-being of people in the exoplanet and disk imaging community, and the possible biases and inappropriate behaviors they experience. The previous publication and this one add to a large and very well documented bibliography on discrimination in scientific fields and in academia, with a majority focused on the topic of gender. The main originality of this article lies in its focus on the professional community (starting from graduate students), whereas the majority of STEM socio-demographic studies tend to concentrate on undergraduate students.

We have focused the survey and our study on different topics: 1) survey participant demographics, 2) visibility and exposure at conferences, 3) recognition through publications and projects, 4) disrespect, 5) inappropriate behaviors, and 6) allyship, which, to our knowledge, is focused on for the first time. These topics have been studied through several perspectives: 1) job position, 2) expatriation status, 3) gender, 4) belonging to another under-represented group in astronomy, and 5) parenthood and more specifically fatherhood. This scope is larger than the 2019 survey, even though there remain limitations with small sample size groups (not statistically relevant and problematic for respondent anonymity) and with asking about some traditionally discriminated groups (as written in Sec. \ref{sec:Demographics}, questions regarding ethnicity or race are forbidden in France). 

Among numerous results, we wish to point out some specific conclusions:

$\bullet$ People from traditionally less-discriminated categories (senior researchers, men...) appear to have participated less in the survey. In addition to biasing the results of this survey, this may indicate a lack of concern for the topics addressed in this paper, which is problematic since such group members are in the best position to serve as allies (see Sec.\ref{sec:Demographics}).

$\bullet$ The well-known leaky pipeline phenomenon was indicated here and shows a correlation between gender and career step. As developed in more focused publications, it could have different sources: women/non-binary people leaving academia more than men, more women/non-binary people arriving recently in the field (see Sec.\ref{sec:Demographics}).

$\bullet$ PhD students tend to have less talks at conferences despite a high need for exposure, since they are in short-term contracts and need collaborations for their next position (see Sec.\ref{sec:Visibility and exposure at conferences}).

$\bullet$ Non-permanent researchers are rarely invited into international conference SOCs. Their inclusion into conference organization would recognize their expertise and their strong and diverse networks, and this would also serve to train early career community members for organizing future conferences (see Sec.\ref{sec:Visibility and exposure at conferences}). However, this also implies allocating time away from their research.

$\bullet$ Postdoctorate researchers, female and non-binary people, and people from groups under-represented in astronomy are biased against in terms of inclusion, specifically in publications and projects (see Sec.\ref{sec:Recognition with publications and projects}).

$\bullet$ More than half of postdoctoral researchers, female and non-binary people, and people from groups underrepresented in astronomy have experienced disrespect within the last $2.5$ years: respectively $52\%$, $52\%$, and $51\%$ (see Sec.\ref{sec:Disrespect}).

$\bullet$ Women and non-binary people are the most exposed to inappropriate behaviors. We also note that early career people are more confronted with inappropriate behaviors: This shows a correlation between being junior (i.e., in a situation of potential professional vulnerability) and being exposed to inappropriate behaviors, which is not acceptable (see Sec.\ref{sec:Inappropriate behaviors}).

$\bullet$ For both disrespect and inappropriate behavior, there is a strong correlation between having experienced it and having witnessed it. People who have experienced such a situation are more watchful for signs of it happening to other people (see Sec.\ref{sec:Disrespect} and \ref{sec:Inappropriate behaviors}).

$\bullet$ Most people consider themselves as allies, with a slight dependency for categories that tend to be impacted by biases (junior researchers, expatriates, female and non-binary people, people from under-represented groups) (see Sec.\ref{sec:Allies}).

$\bullet$ There is a general wish for information about allyship, even among people who identify themselves as non-allies (see Sec.\ref{sec:Allies}). 

It should be notified here that discriminatory and inappropriate behaviors are subject to criminal liability in many countries (e.g., Civil Rights Act of 1964 in the USA, Employment Equality Directive 2000/78/EC in Europe, Equality Act 2010 in the United Kingdom). From these observations and in the different sections of this paper, recommendations have emerged:

\textbf{$\rightarrow$ Opening the organization of conferences for expatriate and non-permanent researchers.}

\textbf{$\rightarrow$ Providing childcare resources to help parents attend conferences, which would be particularly beneficial for mothers.}

\textbf{$\rightarrow$ Being careful about whom to include in publications and projects and particularly discussing their inclusion with postdocs, female and non-binary people, and people from under-represented groups. For example, this can take the form of a code of publishing conduct within consortia.}

\textbf{$\rightarrow$ Spreading or getting information, for instance, through training or talks, to be able to interpret a possible situation of abuse (disrespect, inappropriate behavior, harassment), and evaluating the well-being of colleagues. This information and training could be especially directed at supervisors, faculty researchers, and men, including from the perspective of people from traditionally discriminated groups.}

\textbf{$\rightarrow$ Spreading or getting information, for instance with training or talks, to be able to react properly when confronted with a possible situation of abuse (disrespect, inappropriate behavior, harassment). This could include: appropriate behavior towards the possible victim, people to contact (advisor, human resources, lab equity referees, anti-harassment cell i.e., people external from the institution trained to identify and provide advices in case of harassment), existing procedures, etc. This could target everybody, and particularly people to refer to in case of issues, such as supervisors, men, senior researchers, and others who may be less vulnerable and in a better position to help.}


\textbf{$\rightarrow$ When witnessing a situation of abuse (disrespect, inappropriate behavior, harassment), first always checking with the possible victim how they feel and their needs. It could also help to verify our own interpretation of the situation.}

\textbf{$\rightarrow$ A dedicated training session on existing procedures (in case of disrespect, inappropriate behavior, or abuse) could be offered at the start of the PhD, as junior scientists (including PhD candidates) appear to be the most likely to encounter inappropriate behavior.}

\textbf{$\rightarrow$ Improving the safety of the workplace for everybody, and in particular for early career scientists, to preserve the well-being and overall inclusion and diversity in the field by helping them to stay safely in academia. This includes highlighting the procedures in case of inappropriate behaviors (victim or witness), for instance with training or posters.}

These recommendations can be supported or completed by most of the comments left by respondents in the final open comment question of the survey (question 30). Those responses are summarized below: 

\textbf{$\rightarrow$ At conferences: making pronoun identification available but optional during registration, and clearly specifying where this information will be disclosed.}

\textbf{$\rightarrow$ At conferences: during abstract/poster/talk submission, including IDEA actions. IDEA should be recognized as fundamental when doing science and engineering research and integrated in our daily lives as researchers rather than treated as an unrelated topic.}

\textbf{$\rightarrow$ At conferences: during the program setup, establishing gender quotas (at least $40\%$) and similar for other discriminated groups.}

\textbf{$\rightarrow$ At conferences: having a systematic IDEA talk, for any conference (large or small), more than $20$ minutes long, and with time left for questions and discussions. Different themes can be developed: 1) micro-aggression, disrespect, inappropriate behavior, harassment, 2) allyship and how to react or respond as victim/witness especially in situations with power dynamics, 3) facts and numbers about groups under-represented in astronomy.}

\textbf{$\rightarrow$ At conferences: during breaks, and particularly as a senior researcher, making an effort to spend time with junior researchers and researchers from under-represented communities rather than staying only among permanent researchers.}

\textbf{$\rightarrow$ In projects: giving Principal Investigator (PI)-ship to people from under-represented communities, including women. This can also foster a safe working environment, free from harassment (at institutions and in the field).}

\textbf{$\rightarrow$ In projects: not collaborating with well-known harassers.}

\textbf{$\rightarrow$ In projects and at institutions: systematically opening collaborations, committees, and discussions to junior people and researchers from discriminated groups.}

\textbf{$\rightarrow$ At institutions: providing a clear, easy, and neutral procedure in case of harassment (for instance: externalized from the institution) to avoid conflicts of interest and discouragement.}

An additional comment submitted by a respondent suggested avoiding grouping people into categories. We do believe that this would be the goal of any society or community, but in the short term, it would prevent quantifying issues and it would obscure problematic social behaviors or systemic biases as well as make it harder to prove they are systemic. 

Compared to the 2019 survey, we tried to reduce the biases and limitations of our questions and scope. However, some remain: first, the sample of respondents is quite small which leads to small number statistics for some categories. We set the threshold for quantification to $\geq 7$ people groups so we could include fatherhood as a discrimination factor even if clear trends are complex to evaluate. Second, the respondent panel is biased towards non-permanent researchers and women and non-binary people, since fewer senior people and men answered the survey than estimated attendees from the conference participation rates. In addition, several questions target the feelings or interpretations by people, since some situations cannot be quantified numerically. Furthermore, this conference was the first Spirit of Lyot since the COVID-19 pandemic and lockdown, which could have made it quite exceptional. For instance, it gathered $218$ people vs. $190$ for the 2019 one, and the location was also more central for Europeans who represent a large fraction of the community.

We do recommend organizing another survey at the next Spirit of Lyot to monitor the socio-demographic evolution of our community. Comments from question 30 also encourage the setup of this next survey, with additional recommendations like 1) adding ``I don't remember" options for questions regarding past experiences or witnessing, 2) letting allies and/or permanent researchers take care of the survey to reduce the burden on early career/under-represented groups. We also believe that social scientists as experts in socio-demographic studies should be even more involved to reduce biases of partial questions, bring context, and improve our analysis (see acknowledgments).


\appendix

\section{Comparison among expertises}
\label{sec:AppendixB}

This appendix aims at comparing a few outcomes regarding the expertises. Since only 4 respondents have declared Theory as their main research interest, the results later on focus on Instrumentation and Observations only. 

\begin{figure}
   \begin{center}
   \begin{tabular}{ccc}
   \includegraphics[width=5.7cm]{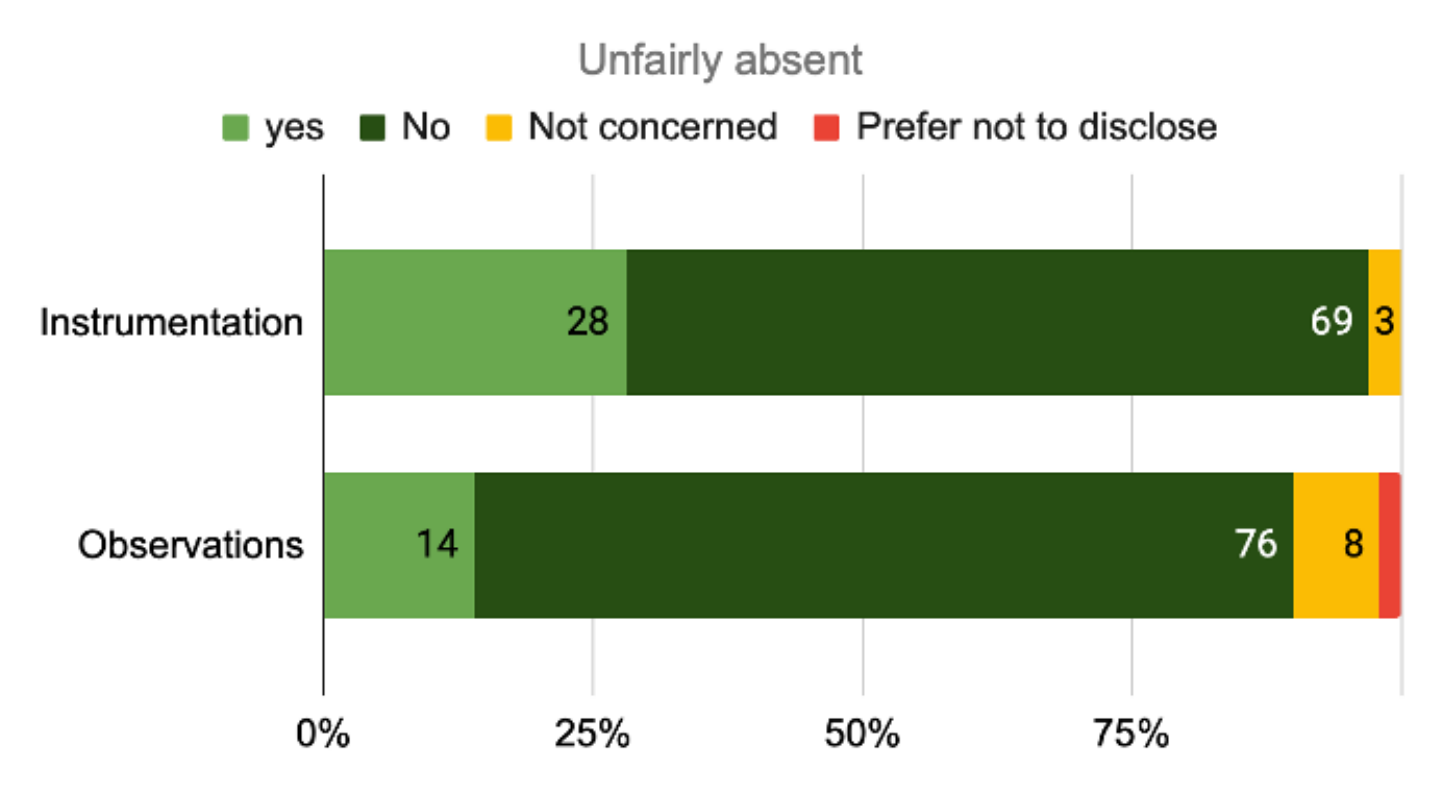} & \includegraphics[width=5cm]{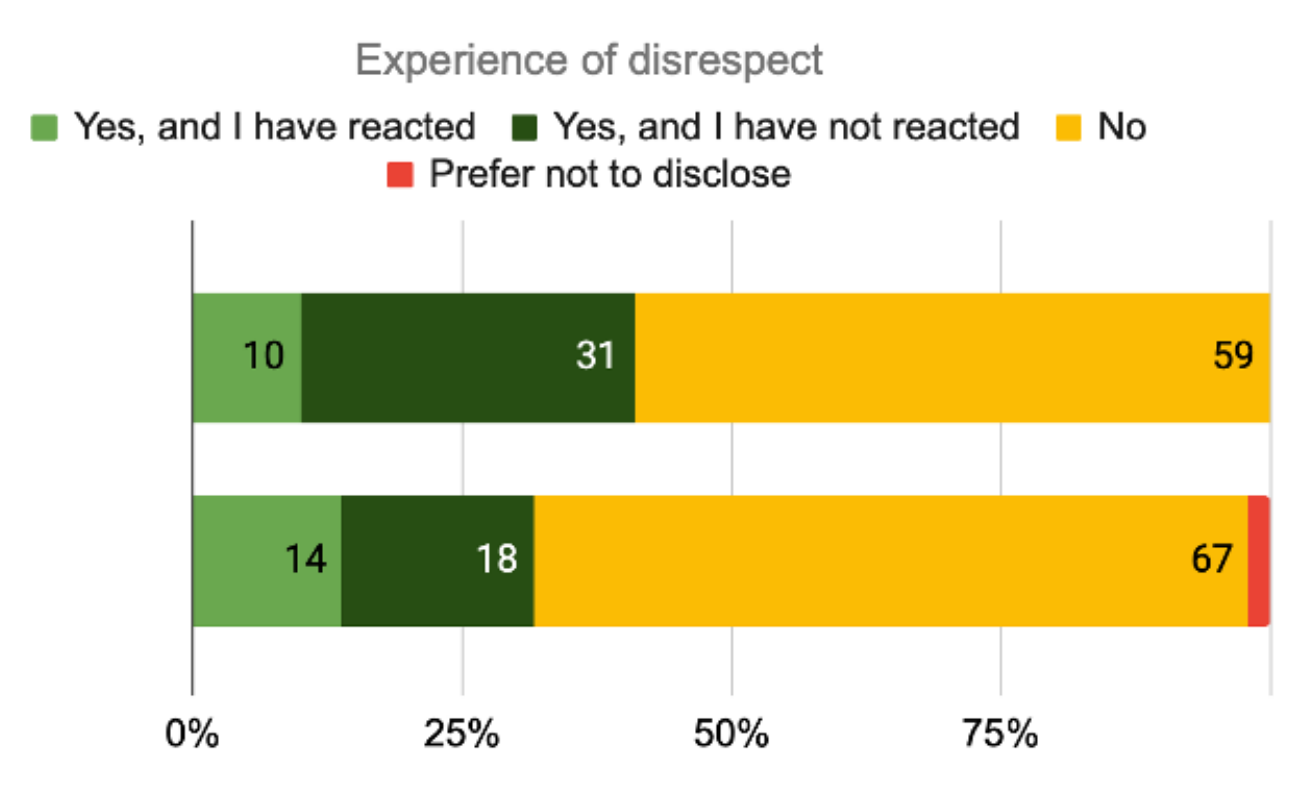} & \includegraphics[width=5cm]{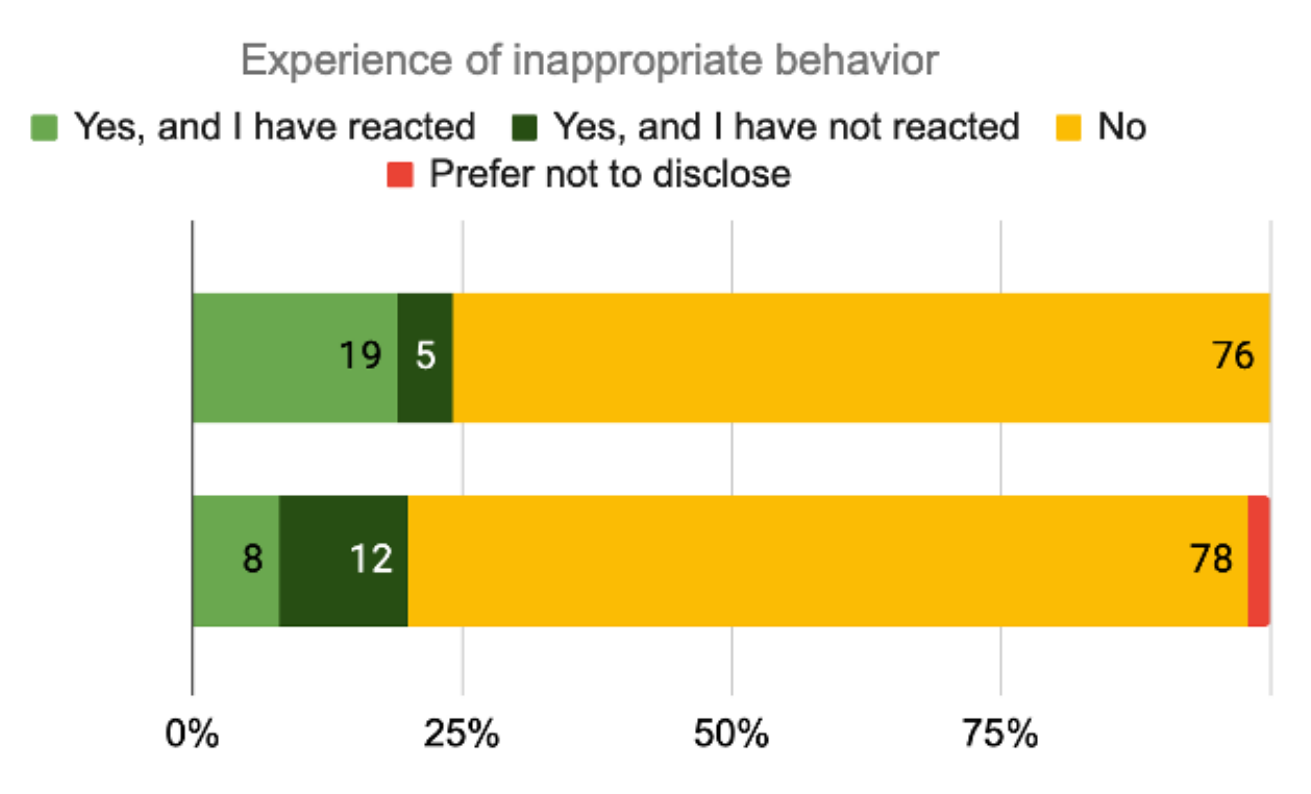} \\
   \end{tabular}
   \end{center}
   \caption
   { \label{fig:Annexe} 
Outcome regarding the research expertise: (left) percentages of respondents, per category, who have felt unfairly absent from a list of co-authors, (center) percentages of respondents, per category, who declare having experienced disrespect (right) percentages of respondents, per category, who declare having experienced a situation of inappropriate behavior.} 
   \end{figure}

In Figure \ref{fig:Annexe}, three outcomes are compared: the percentages of people having felt unfairly absent from co-authorship, having experiences disrespect, and having experiences a situation of inappropriate behavior:

$\bullet$ Concerning recognition, experts in instrumentation declare twice more than experts in observations having felt unfairly left aside from co-authorship. This result can support the set up of a code of publishing conduct in instrumental research. 

$\bullet$ Concerning disrespect and inappropriate behaviors, Instrumentation appears slightly less safe than Observations with $41\%$ of respondents in Instrumentation having experienced disrespect versus $32\%$ in Observations. Regarding situations of inappropriate behavior, no clear trend appears with $24\%$ of respondents in Instrumentation having experienced it versus $20\%$ in Observations.

\section{Survey}
\label{sec:AppendixA}

\quad \textbf{Demographics:}

1) Choose the option that best describes your gender: \\
\textit{Female / Male / Non binary / Other / Prefer not to disclose}

2) Do you consider yourself as an expatriate?\\
\textit{Yes / No / Prefer not to disclose}

3) What continent do you currently work in?\\
\textit{Africa / Antarctica / Asia / Europe / North America / Oceania / South America / Prefer not to disclose}

4) Do you consider yourself as part of an under-represented group in astronomy (in terms of ethnicity, disability, or sexual orientation)?\\
\textit{Yes / No / Prefer not to disclose}

5) Do you have children?\\
\textit{Yes / No / Prefer not to disclose}

6) What kind of position do you occupy?\\
\textit{Intern / PhD student / Postdoc / Faculty / Other / Prefer not to disclose}

7) What is your expertise? (several choices)\\
\textit{Instrumentation / Observation / Theory / Prefer not to disclose}

\textbf{Visibility/Conferences:}

8) How many international conferences did you attend (in person or virtually) since January 2020?\\
\textit{0 / 1 / 2 / 3 / 4 / 5 or more / Not concerned / Prefer not to disclose}

9) At these conferences, how many talks have you given in total?\\
\textit{0 / 1 / 2 / 3 / 4 / 5 or more / Not concerned / Prefer not to disclose}

10) Have you been invited to join the SOC of an international conference since January 2020?\\
\textit{Yes / No / Not concerned / Prefer not to disclose}

11) How many seminars have you been invited to give (in person or virtually) since January 2020?\\
\textit{0 / 1 / 2 / 3 / 4 / 5 or more / Not concerned / Prefer not to disclose}

\textbf{Recognition/Papers and projects:}

12) How many peer-reviewed articles have you published as first author since January 2020?\\
\textit{0 / 1 / 2 / 3 / 4 / 5 or more / Not concerned / Prefer not to disclose}

13) How many articles have you been asked to review since January 2020?\\
\textit{0 / 1 / 2 / 3 / 4 / 5 or more / Not concerned / Prefer not to disclose}

14) Have you ever felt unfairly absent from the list of co-authors of a publication since January 2020?\\
\textit{Yes / No / Not concerned / Prefer not to disclose}

15) Have you ever felt unfairly present in the list of co-authors of a publication since January 2020?\\
\textit{Yes / No / Not concerned / Prefer not to disclose}

16) Have you felt left aside from a project or publication you deserved to be part of (implication or leadership) since January 2020?\\
\textit{Yes / No / Not concerned / Prefer not to disclose}

\textbf{Disrespect:}

By disrespect, we mean light behaviors that can make you or somebody uncomfortable, for instance cutting off somebody while they are talking, talking over one other project without their consent, downplaying someone’s idea, expecting social role from younger people or women (taking notes, planning social events...). More reprehensible behaviors will be the purpose of the next section.

17) Since January 2020, have you experienced, at work or at a conference (online or in person), a situation of disrespect?\\
\textit{Yes and I have reacted / Yes and I have not reacted / No / Prefer not to disclose}

18) If ``yes" at question 18, have you received support from other people?\\
\textit{Yes / No, no one noticed / No, they would not understand / Prefer not to disclose}

19) Since January 2020, have you witnessed, at work or at a conference (online or in person), a situation of inappropriate behavior?\\
\textit{Yes and I have reacted / Yes and I have not reacted / No / Prefer not to disclose}

20) If 'yes and I have responded' at question 19, can you provide examples of actions you have taken?

21) If 'Yes and I have not reacted' in question 19, would you like to share why you did not react?

\textbf{Inappropriate behaviors:}

By inappropriate behaviors, we mean any social behavior that can make you uncomfortable even if it is not necessarily legally reprehensible. Here are a few specific examples of inappropriate behaviors: condescending remarks, discriminating behavior based on ethnicity, inappropriate jokes, racist jokes, staring at, sexual remarks or questions at a work environment, disrespect based on one's culture and identity, ignoring or excluding somebody during a meeting, preventing somebody from attending meetings, remarks on parental leaves, sexual harassment, bullying...

22) Since January 2020, have you experienced, at work or at a conference (online or in person), a situation of inappropriate behavior?\\
\textit{Yes and I have reacted / Yes and I have not reacted / No / Prefer not to disclose}

23) If yes at question 18, have you received support from other people?\\
\textit{Yes / No, no one noticed / No, they would not understand / Prefer not to disclose}

24) Since January 2020, have you witnessed, at work or at a conference (online or in person), a situation of inappropriate behavior?\\
\textit{Yes and I have reacted / Yes and I have not reacted / No / Prefer not to disclose}

25) If 'yes and I have responded' in question 24, can you provide examples of actions you have taken?

26) If 'yes and I have not reacted' at Question 24, would you like to share why you did not react?

\textbf{Allies:}

An ally is somebody who 1) acknowledges the difficulties faced by underrepresented or discriminated groups they do not belong to and 2) supports them when necessary or requested.

27) Do you see yourself as an ally of underrepresented groups in your team?\\
\textit{Yes / No / Prefer not to disclose}

28) Would you like talks or information about how to be an ally of underrepresented groups?\\
\textit{Yes / No / Prefer not to disclose}

29) Are you aware of the procedure in the event of harassment in your institution?\\
\textit{Yes / No / Prefer not to disclose}

\textbf{Comments and feedbacks:}

30) Are there any comments or feedback that you would like to share with us? e.g., are there behaviors (specific to conferences or not) you would like to point out? e.g., would you like conferences to schedule more talks about EDI (Equity, Diversity, Inclusion) and awareness on social behaviors in the future? e.g. if you were given power to change at least one fact about the high-contrast imaging community, what would it be? ...

\acknowledgments

The authors wish to thank the Spirit of Lyot 2022 committees for requesting this survey to be performed and for providing support in its setup. They are also deeply grateful to Yazmin Gonzalez for her feedback that has greatly contributed to improving the quality of the paper. The authors also thank the respondents for their participation in this survey and their trust. They are well aware that answering such a survey is not easy, both in terms of topic to discuss, experience to share, and fear for one's anonymity to be released. They hope that the choices they made to present the survey results meet the expectations and trust of the participants. Eventually, the authors also wish to thank the editor and both reviewers of this paper for their careful read as well as interesting and helpful remarks.

\bibliography{main}{}
\bibliographystyle{aasjournal}

\end{document}